%% file: shinglesetal2015.tex
\documentclass[useAMS,usenatbib]{mn2e}

\usepackage{graphicx}
\usepackage{amsmath}
\usepackage{amssymb}
\usepackage{microtype}
\usepackage{multicol,multirow}
\usepackage{verbatim}
\usepackage{xcolor}

\newcommand{\Mdot}{\ensuremath{\dot{\mathrm{M}}}}
\newcommand{\Msun}{\ensuremath{\mathrm{M}_{\sun}}}
\newcommand{\Lsub}[1]{\ensuremath{\mathrm{L}_\mathrm{#1}}}

\newcommand{\M}[1]{\ensuremath{\mathrm{M}_\mathrm{#1}}}
\newcommand{\Mtwo}[2]{\ensuremath{\mathrm{M}_\mathrm{#1}^\mathrm{#2}}}
\newcommand{\Tsub}[1]{\ensuremath{\mathrm{T}_\mathrm{#1}}}
\newcommand{\Ttwo}[2]{\ensuremath{\mathrm{T}_\mathrm{#1}^\mathrm{#2}}}
\newcommand{\tausub}[1]{\ensuremath{\tau_\mathrm{#1}}}
\newcommand{\Mc}{\M{C}}

\newcommand{\Mtdu}{\ensuremath{\Mtwo{TDU}{tot}}}
\newcommand{\iso}[2]{\hbox{${}^{#1}{\text{#2}}$}}

\title[He-rich AGB evolution and nucleosynthesis]{Evolution and nucleosynthesis of helium-rich asymptotic giant branch models}
\author[Shingles et al.]{Luke J. Shingles$^{1}$\thanks{Email:
luke.shingles@anu.edu.au}, Carolyn L. Doherty$^{2}$, Amanda I. Karakas$^{1}$, Richard J. Stancliffe$^{3}$,\newauthor John C. Lattanzio$^{2}$, and Maria Lugaro$^{4}$\\
$^{1}$Research School of Astronomy and Astrophysics, Australian National University, Canberra, ACT 2611, Australia\\
$^{2}$Monash Centre for Astrophysics (MoCA), School of Physics and Astronomy, Monash University, Victoria 3800, Australia\\
$^{3}$Argelander-Institut f\"ur Astronomie, University of Bonn, Auf dem H\"ugel 71, 53121 Bonn, Germany\\
$^{4}$Konkoly Observatory, Hungarian Academy of Sciences, PO Box 67, H-1525 Budapest, Hungary}

\date{Accepted 2015 July 2.  Received 2015 July 2; in original form 2015 April 29}
\pubyear{2015}
\begin{document}
\label{firstpage}
\pagerange{\pageref{firstpage}--\pageref{lastpage}}
\maketitle

\begin{abstract}
There is now strong evidence that some stars have been born with He mass fractions as high as $Y\approx 0.40$ (e.g., in $\omega$ Centauri). However, the advanced evolution, chemical yields, and final fates of He-rich stars are largely unexplored. We investigate the consequences of He-enhancement on the evolution and nucleosynthesis of intermediate-mass asymptotic giant branch (AGB) models of 3, 4, 5, and 6 \Msun\ with a metallicity of $Z = 0.0006$ ([Fe/H] $\approx -1.4$). We compare models with He-enhanced compositions ($Y=0.30, 0.35, 0.40$) to those with primordial He ($Y=0.24$). We find that the minimum initial mass for C burning and super-AGB stars with CO(Ne) or ONe cores decreases from above our highest mass of $6\ \Msun$ to $\sim 4$--$5\ \Msun$ with $Y=0.40$. We also model the production of trans-Fe elements via the \textit{slow} neutron-capture process (\textit{s}-process). He-enhancement substantially reduces the third dredge-up efficiency and the stellar yields of \textit{s}-process elements (e.g., 90\% less Ba for 6 \Msun, $Y=0.40$). An exception occurs for 3 \Msun, where the near-doubling in the number of thermal pulses with $Y=0.40$ leads to $\sim50\%$ higher yields of Ba-peak elements and Pb if the \iso{13}{C} neutron source is included. However, the thinner intershell and increased temperatures at the base of the convective envelope with $Y=0.40$ probably inhibit the \iso{13}{C} neutron source at this mass. Future chemical evolution models with our yields might explain the evolution of \textit{s}-process elements among He-rich stars in $\omega$ Centauri.
\end{abstract}


\section{Introduction}\label{sec:intro}
Stars that evolve through the asymptotic giant branch (AGB) play a crucial role in the chemical evolution of stellar populations and galaxies. These stars experience a complex sequence of He-shell instabilities, mixing, and mass loss processes that combine to eject material that has undergone H and He burning, and enrichment in heavy elements produced by the \textit{slow} neutron-capture process (\textit{s}-process). Understanding the chemical contribution by stars to the interstellar medium is a prerequisite for building models of chemical evolution \citep[e.g.,][]{Chiappini:2001ds,Kobayashi:2011hj}.

The most important parameter governing stellar evolution is the initial mass, with a secondary role played by chemical composition. Chemical composition typically refers to the mass fraction of metals ($Z$), but the mass fraction of He ($Y$) also has major consequences for stellar evolution. However, the stellar evolution and nucleosynthesis that take place during the AGB phase are relatively unexplored for He-rich initial compositions. For a detailed introduction to AGB evolution and nucleosynthesis with normal He content, we refer the reader to the reviews by \citet{Herwig:2005jn} and \citet{Karakas:2014jt}.

There is strong evidence that some stars have been born with substantial He enrichments above the primordial He mass fraction of $Y\approx 0.24$ predicted from big bang nucleosynthesis. The most direct method to measure He abundances is to use spectroscopy, however the application to He is severely hindered by the lack of He lines in cool stars, and the effects of gravitational settling in hot stars. Still, direct detections of He have been made, such as those by \citet{Pasquini:2011kf}, who used the He I 10830 {\AA} line in two stars with different Na and O abundances in NGC 2808 to derive a difference of $\Delta Y \geq 0.17$. The finding of He-enhanced second-generation stars in NGC 2808 is also supported by \citet{Marino:2014fj} with a sample of 96 horizontal-branch stars, although they report a smaller average enhancement of $\Delta Y = 0.09 \pm 0.06$. The same He I 10830 {\AA} line was also used by \citet{Dupree:2013cb} to infer a variation of $\Delta Y \geq 0.17$ between red giants in $\omega$ Centauri.

A second, less-direct line of evidence for He-rich stars is the requirement of high He abundances to reconcile theoretical isochrones with the photometry of clusters in colour-magnitude diagrams (CMDs). For example, the split main sequence of $\omega$ Centauri can be explained by multiple populations that vary in both metallicity and He abundance \citep{Bedin:2004hn,Norris:2004jg,Piotto:2005ii}. \citet{King:2012bf} report a best fit to the blue main sequence with a He abundance of $Y=0.39 \pm 0.02$. The high He abundance also persists when additional features of the CMD are incorporated in a simultaneous fit, such as the work of \citet{Joo:2013dr}, who infer $Y$ of up to $0.39\pm 0.02$ for $\omega$ Centauri and up to $0.32\pm 0.04$ for M22.

With relatively few theoretical studies of the detailed AGB evolution and nucleosynthesis of He-rich models \citep{Campbell:2013bd,Charbonnel:2013jt,Karakas:2014ja,Karakas:2014jv}, the effects of high He abundances on chemical evolution are not well understood. An important set of observational clues are the \textit{s}-process abundances in $\omega$ Centauri. In this cluster, [Ba/Fe]\footnote{We use the standard spectroscopic notation, [A/B] $=\log_{10}(N_A/N_B) - \log_{10}(N_A^{\sun} / N_B^{\sun})$, where $N_A$ and $N_B$ are abundances by number and $\sun$ denotes the solar abundance.} increases with metallicity up to about [Fe/H] $=-1.4$, above which [Ba/Fe] remains roughly constant \citep{Norris:1995kc}. Interestingly, this plateau occurs near the metallicity of the most He-rich population \citep{Joo:2013dr}, and one potential explanation is that Ba yields are lower in He-rich stars. Chemical evolution might play out similarly in other clusters such as M22, which follows the same rise of [Ba/Fe] \citep{DaCosta:2011jh} but where a truncation of the chemical evolution at a metallicity below [Fe/H]$=-1.4$ prevents the possibility of a similar \textit{s}-process plateau.

\citet*{Karakas:2014ja} (hereafter KMN14) present stellar yields of He-enhanced 1.7 \Msun\ and 2.36 \Msun\ AGB models with a metallicity of [Fe/H] $\approx-1.4$. This metallicity is appropriate for the most metal-poor of the He-rich populations in $\omega$ Centauri \citep{Joo:2013dr}. Among the findings of KMN14 is a reduction in the third dredge-up mass with increasing initial He abundance. Largely as a result of the less efficient dredge up, their Ba yields decrease by roughly 50 per cent with $Y=0.40$ in comparison with primordial-He models at the same initial mass.

In this work, we extend the study of He-enhanced AGB nucleosynthesis by KMN14 with new intermediate-mass models of 3, 4, 5, and 6 \Msun\ at the same metallicity of $Z=0.0006$ and with similar input physics. For our primordial He models, we make a comparison with the models of \citet*{Straniero:2014jk} (hereafter SCP14), which have $Y=0.245$, $Z=0.0003$ (before $\alpha$-enhancement, [Fe/H] $=-1.7$), and [$\alpha$/Fe] $=0.5$. We also compare our evolutionary models with the primordial-He models of \citet{Ventura:2009ha} (hereafter VD09), which have $Z=0.0006$ and [$\alpha$/Fe] $=0.4$. The models of VD09 are calculated with the Full Spectrum of Turbulence \citep[FST,][]{Canuto:1991bj} treatment of convective mixing and energy transport rather than the more common Mixing Length Theory \citep[MLT,][]{Biermann:1948ul}.

The stellar yields of He-rich models, including those presented in this study will build a foundation for future work in understanding the chemical evolution of He-rich environments. Our comparison with other models in the literature allows us to estimate how sensitive our predictions for He-rich stellar evolution and yields are to uncertain modelling assumptions.

In Section \ref{sec:method} we describe our computational method and stellar modelling assumptions. In Section \ref{sec:evolutionmodels} we present our stellar evolutionary models. In Section \ref{sec:nucleosynthesisyields} we present the nucleosynthesis models and explore the impact of He-enhancement on the stellar yields. In Section \ref{sec:discussion} we discuss our results in the context of other studies in the literature and consider the implications for He-rich chemical evolution.

\section{Computational Method}\label{sec:method}
We present evolutionary sequences for initial masses of 3, 4, 5, and 6 \Msun\ with a global metallicity of $Z=0.0006$ ([Fe/H] $\approx-1.4$ and [$\alpha$/Fe]$=0.0$), and initial He mass fractions of $Y=0.24, 0.30, 0.35,$ and $0.40$.

We use the same input physics and stellar evolutionary code as KMN14 (the \textsc{Mount Stromlo Stellar Structure Program}), except that we include the opacity treatment of \citet{Fishlock:2014hn}, which accounts for changes to the C abundance and the C/O ratio at the surface. This opacity treatment improves the accuracy for models that experience hot bottom burning (discussed in Section \ref{sec:hbb}). At low-temperatures ($T < 10,000$ K), we use the molecular opacity tables from AESOPUS \citep{Marigo:2009gz}. At higher temperatures, we use OPAL tables \citep{Iglesias:1996dp} that have been updated to a \citet{Lodders:2003bf} scaled-solar composition.

The initial abundances of the models are scaled solar \citep{Asplund:2009eu} with a global metallicity of $Z=0.0006$. With $Z$ fixed and a chosen value of $Y$, the H mass fraction ($X$) is determined by $X=1-Y-Z$. A consequence of keeping the mass fraction of metals (including Fe) constant and exchanging mass between H and He is that the initial ratio of Fe to H varies slightly between our models of different He abundance. Specifically, the initial [Fe/H] is $-1.41$ in the models for $Y=0.24$, and $-1.31$ in the models for $Y=0.40$.

The search for stable convective borders uses the method of \citet{Lattanzio:1986cz} \citep[see also][]{Frost:1996fu}, and we assume no other form of extra mixing at convective boundaries. The convective velocities are calculated using MLT with a mixing length parameter of $\alpha=1.86$. Although the constancy of $\alpha$ is a standard assumption, empirical and theoretical studies have found a variation of this parameter with stellar evolution \citep{Lebzelter:2007ho,Magic:2014vb}. Larger values of $\alpha$ can increase the depth of the third dredge-up \citep[e.g.,][]{Boothroyd:1988jl}, which in turn typically increases the yields of C+N+O and \textit{s}-process elements \citep{Cristallo:2009kn,Cristallo:2011fz}.

We do not include mass loss on the first red giant branch (RGB). This assumption is reasonable (in the context of stellar yield predictions) given recent results using Kepler data, which indicate that the total mass loss during the RGB is relatively small \citep[$\Delta\mathrm{M}=0.09\pm 0.03\pm 0.04\ \Msun$,][]{Miglio:2012dm}, at least near solar-metallicity. Many of our models do not reach the RGB and for the models that do, the inclusion of mass loss of the this order would have a minimal impact on our stellar yield predictions, which are subject to much larger uncertainties related to mass loss during the AGB phase \citep{Stancliffe:2007er}. For mass loss on the AGB, we use the prescription given by Equations 1 to 4 of \citet{Vassiliadis:1993jk}, which includes a superwind phase beginning at pulsation periods above approximately 500 days.

The output of the stellar evolution code (including the temperatures, densities, convective boundaries, and convective velocities) is used as input to a post-process code that performs the detailed nucleosynthesis calculations with a large network of 320 species up to Po. The two-pass method used in this work is different to the models of SCP14 calculated with the \textsc{FUNS} code \citep[a descendant of \textsc{FRANEC},][]{Chieffi:1989dp}, which solves the stellar structure equations simultaneously with the full network of nuclear reactions.

Another difference is that our post-processing code solves simultaneously the abundance changes wrought by mixing and burning. To mix the convective regions in our nucleosynthesis models, we use the two-stream, time-dependent method devised by \citet{Cannon:1993te}. An upward and a downward-moving stream are treated separately, with the mass-flow rate at each mass shell calculated from the convective velocities in the evolutionary model. It is assumed that the material from an upward moving cell below flows into the upward moving cell above, and likewise the material in a downward moving cell flows into the downward moving cell below, as well as horizontally mixing from one stream to the other.

\subsection{The inclusion of \iso{13}{C} pockets}\label{sec:pmz}
For stars with initial masses $\lesssim 3\ \Msun$, the dominant neutron source driving \textit{s}-process nucleosynthesis is the $\iso{13}{C}(\alpha,n)\iso{16}{O}$ reaction \citep{Busso:1999ig}, while the $\iso{22}{Ne}(\alpha,n)\iso{25}{Mg}$ reaction becomes more important at higher initial masses \citep{GarciaHernandez:2013ko}. For our 3 \Msun\ models which are near the mass of this transition, the treatment of \iso{13}{C}-pocket formation is crucial for making accurate predictions of the stellar yields and surface abundances of neutron-capture elements.

In current theoretical models, the $\iso{13}{C}(\alpha,n)\iso{16}{O}$ reaction occurs in a thin layer of \iso{13}{C} that results from the partial mixing of protons from the base of the envelope into the \iso{12}{C}-rich intershell region during third dredge-up \citep{Straniero:1995ed,Gallino:1998eg}. The protons then are captured by \iso{12}{C} where they activate the CN-cycle reactions $\iso{12}{C}(p,\gamma)\iso{13}{N}(\beta^+\nu)\iso{13}{C}$. The number of protons mixed into the region must be low because further proton capture results in \iso{14}{N}, which efficiently absorb free neutrons. The physical mechanism that produces the partially mixed zone (PMZ) has not been conclusively identified, but plausible candidates include convective-boundary mixing \citep{Herwig:2000ua,Cristallo:2009kn}, rotational mixing \citep{Langer:1999tj,Herwig:2001vb}, gravity-wave driven mixing \citep{Denissenkov:2003gx}, and semiconvection \citep{Iben:1982cv,Hollowell:1989bd}. Further, the important effect of rotationally induced mixing on the neutron release in the \iso{13}{C} pocket has been investigated by \citet{Herwig:2003du}, \citet{Siess:2004eh}, and \citet{Piersanti:2013dh}.

The nature of the transition away from \iso{13}{C} pockets with increasing initial mass is also highly uncertain, although it is very likely connected with the shrinking of the He-rich intershell and the higher temperatures at the base of the convective envelope with larger core masses.

For example, the models of SCP14 assume that the convective velocity declines exponentially with distance beyond the formal convective boundary, with a free parameter describing the length scale of the velocity decline. This method is applied only to convective boundaries for which the velocity is discontinuous, which occurs at the base of the envelope during third dredge-up episodes \citep{Cristallo:2009kn}. The convective velocity is incorporated into a non-diffusive mixing scheme \citep{Straniero:2006do} and results in \iso{13}{C} pockets that smoothly decrease in size (and neutron production) with increasing initial mass.

\citet{Goriely:2004gw} show that a diffusive treatment of convective-boundary mixing can reduce the amount of neutron production by \iso{13}{C} pockets (due to enhanced production of \iso{14}{N}) when the third dredge-up takes place with temperatures above 40 MK, and can totally inhibit the \textit{s}-process with temperatures above 70 MK. As shown in Table \ref{tbl:structureagb}, the maximum temperatures at the base of the envelope during third dredge-up for our 3 \Msun\ models with $Y$ between 0.24 and 0.35 occupy this transition range of temperatures. For the 3 \Msun\ model with $Y=0.40$ and the higher-mass models with all He abundances, temperatures above 70 MK during third dredge-up would likely prevent a significant $s$-process production by \iso{13}{C} pockets. For this reason, a PMZ leading to the formation of a \iso{13}{C} pocket is not included in our models with $M \geq 4\ \Msun$. For the initial mass of 3 \Msun, we construct nucleosynthesis models with and without a PMZ at each He abundance.

In the absence of a deep physical understanding of the PMZ, we apply the simple parameterised treatment used by \citet{Lugaro:2004en} \citep[similar to][]{Goriely:2000vs} in our nucleosynthesis post-processing models. For models that include a PMZ, we insert an exponential (in mass coordinate) profile of protons below the envelope at the deepest extent of each third dredge-up episode. The proton abundance starts at the envelope value and decreases by a factor of $10^{4}$ through a depth given by the parameter $\M{pmz}$. The value of $\M{pmz}$ is uncertain, although a variety of constraints have been derived from observations of AGB stars \citep{Abia:2002kh}, C-enhanced metal poor stars \citep{Izzard:2009hg,Bisterzo:2012bf,Lugaro:2012ht}, planetary nebulae \citep{Shingles:2013kg,Miszalski:2013gi}, and post-AGB stars \citep{BonacicMarinovic:2007jl,DeSmedt:2012dp}. KMN14 presented 1.7 \Msun\ and 2.36 \Msun\ models with a range of PMZ sizes, but for comparison we only use their results for a PMZ of $10^{-3}\ \Msun$. Except where stated otherwise, our 3\ \Msun\ nucleosynthesis models include a PMZ with $\M{pmz}=10^{-3}\ \Msun$.

\section{Stellar Evolution Models}\label{sec:evolutionmodels}
\begin{table*}
 \begin{minipage}{170mm}
 \caption{Evolutionary lifetimes, first and second dredge-up masses, and final model properties. We include the initial stellar mass ($\M{ini}$), the initial He mass fraction ($Y$), the core H-burning lifetime ($\tausub{ms}$), the core He-burning lifetime ($\tausub{coreHe}$), the AGB lifetime measured from the exhaustion of core He until the end of calculations plus the estimated time to lose the remaining envelope ($\tausub{agb}$), the total stellar lifetime ($\tausub{stellar}$), the innermost mass layer reached by first and second dredge-up ($\M{FDU}$ and $\M{SDU}$), the final stellar mass ($\Mtwo{}{final}$), the final core mass ($\Mtwo{C}{final}$), the final envelope mass ($\Mtwo{env}{final}$), the final mass-loss rate ($\Mdot^\mathrm{final}$), and the class of white dwarf remnant formed. Final refers the state at the end of our calculations, which cease due to convergence problems before the envelope has been fully ejected.}
 \label{tbl:structure}
 \input{table1.tex}
 \end{minipage}
\end{table*}

\begin{table*}
 \begin{minipage}{180mm}
 \caption{Structural properties relevant to the thermally pulsing AGB phase. We include the initial stellar mass ($\M{ini}$), the initial He mass fraction ($Y$), the H-exhausted core mass at the first thermal pulse ($\Mtwo{c}{TP1}$), the number of calculated thermal pulses (TPs), the dredge-up efficiency $\lambda$ averaged over all TPs ($\left<\lambda\right>$), the maximum dredge-up efficiency ($\lambda_\mathrm{max}$), the total TDU mass ($\Mtwo{TDU}{tot}$), the maximum extent of the pulse-driven convective zone ($\Mtwo{pdcz}{max}$), the maximum temperature in the He-burning shell ($\Ttwo{He-shell}{max}$), the maximum temperature in the H-burning shell $(\Ttwo{H-shell}{max}$), the maximum temperature at the base of the convective envelope during dredge-up or during interpulse H-burning ($\Ttwo{bce}{max,dup}$ and $\Ttwo{bce}{max,ip}$), the average interpulse period ($\left<\tausub{ip}\right>$), and the total duration for which $\Tsub{bce}>50$\,MK ($t_\mathrm{hbb}$).}
 \label{tbl:structureagb}
 \input{table2.tex}
 \end{minipage}
\end{table*}

The evolutionary properties of our models and their dependencies on the initial He abundance are reported in Tables \ref{tbl:structure} and \ref{tbl:structureagb} and discussed below.

\subsection{Pre-AGB evolution and stellar lifetimes}
\begin{figure}
 \begin{center}\includegraphics[width=\columnwidth]{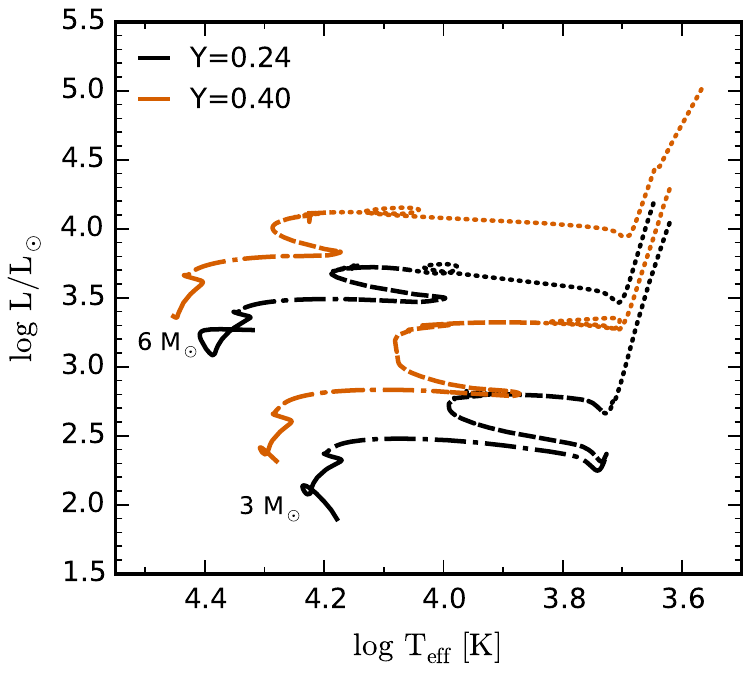}\end{center}
 \caption{Evolutionary Hertzsprung--Russell tracks of 3 and 6 \Msun\ models at $Y=0.24$ (black lines) and $Y=0.40$ (orange lines) from the main sequence to the beginning of the thermally pulsing AGB phase. Line styles indicate the nuclear burning stages: core H burning (solid), shell H burning (dot dashed), core He burning (dashed), and shell He and H burning (dotted).\label{fig:hrdiag}}
\end{figure}

Figure \ref{fig:hrdiag} shows the evolutionary tracks of the 3 \Msun\ and 6 \Msun\ models with $Y=0.24$ and $0.40$ in a Hertzsprung-Russell diagram. In He-enhanced models, the initial compositions have a higher mean molecular weight (as H is exchanged for He), which leads to higher luminosities during the core H-burning phase \citep{Roeser:1975ws}. Compared with models for $Y=0.24$, the models for $Y=0.40$ are roughly twice as luminous during the main sequence. They are also more luminous at each corresponding stage of evolution due to the larger core masses of He-enhanced models.

As shown by the H-burning time-scales and total stellar lifetimes in Table \ref{tbl:structure}, the increased H-burning rate and the reduced H fuel available in He-enhanced models lead to much shorter core H-burning lifetimes and shorter stellar lifetimes overall. This has important implications for chemical evolution, as He-enhanced stars will process and eject their enriched material much earlier than primordial-He stars at a given initial stellar mass. With $Y=0.40$, the stellar lifetimes of the models are reduced by roughly 50 per cent compared to models with primordial He abundance. The shortening of stellar lifetimes by a similar factor with $Y=0.40$ is also found by \citet{Charbonnel:2013jt} for low-mass models between 0.66 \Msun\ and 0.82 \Msun\ at [Fe/H] $=-1.56$, and KMN14 for 1.7 \Msun\ and 2.36 \Msun\ models with the same metallicity of $Z=0.0006$ as the models presented here.

\subsection{The first and second dredge-up}
After the exhaustion of H in the core, the onset of H burning in a shell around the core causes the outer layers to expand and cool as a star begins to ascend the red giant branch. At this stage, the envelope convection zone can move inwards in mass and dredge up the products of partial H burning, which increases the surface abundances of \iso{13}{C} and \iso{14}{N} and decreases the abundance of \iso{12}{C}. Dredge-up at this evolutionary stage is known as first dredge-up (FDU).

Table \ref{tbl:structure} shows the innermost mass layer reached by the convective envelope during FDU for models in which FDU occurs. The depth of FDU in the 3 \Msun\ models decreases with increasing He abundance up to $Y=0.40$, for which FDU does not take place. This is because these stars ignite He in their cores before their envelopes expand sufficiently to reach the first giant branch (Figure \ref{fig:hrdiag}). We find no FDU for initial masses of 4 \Msun\ and above at any He abundance. For the $Y=0.24$ models, the lack of FDU with initial masses of 4 \Msun\ and higher is consistent with the slightly higher-metallicity ($Z=0.001$) models of \citet{Fishlock:2014jq}, for which FDU ceases between initial masses of 3.5 and 4.0 \Msun.

Following the core He-burning phase, the stars change to a dual shell-burning structure and begin their ascent of the AGB. As the He-shell is ignited and the H-shell becomes extinguished for the first time, the inner boundary of the convective envelope moves inwards in mass and dredges up the products of complete H burning, which increases the surface abundances of \iso{4}{He} and \iso{14}{N}. This is the second dredge-up (SDU).

Table \ref{tbl:structure} shows the innermost mass later reached by the SDU for each of our models. The SDU is deeper than the FDU in every model with FDU, and the depth of the SDU decreases with increasing He abundance. However, due to the increased core mass in the 6 \Msun\ model with $Y=0.40$, a ``corrosive SDU'' takes place \citep{GilPons:2013ch,Doherty:2014hx}. In this model, the inner edge of the convective envelope reaches below the top of the CO core, which dredges up C and O to the surface.

\subsection{The thermally pulsing AGB and third dredge-up}\label{sec:tpagb}
During the AGB phase, thermal pulse cycles are driven by thermal instabilities of the thin He-burning shell \citep{Iben:1975js,Herwig:2005jn}. The rapid flash burning of the He-shell releases an enormous amount of energy and causes a pulse-driven convective zone to form, which homogenises the composition of the He-intershell. As the energy moves upwards and expands the star, the H-burning shell is effectively extinguished. With the H-burning barrier removed, the envelope convection zone can penetrate into the He-intershell in an event known as a third dredge-up (TDU) episode \citep{Iben:1975js,Sackmann:1980ed}.

\begin{figure*}
 \begin{center}\includegraphics[width=\textwidth]{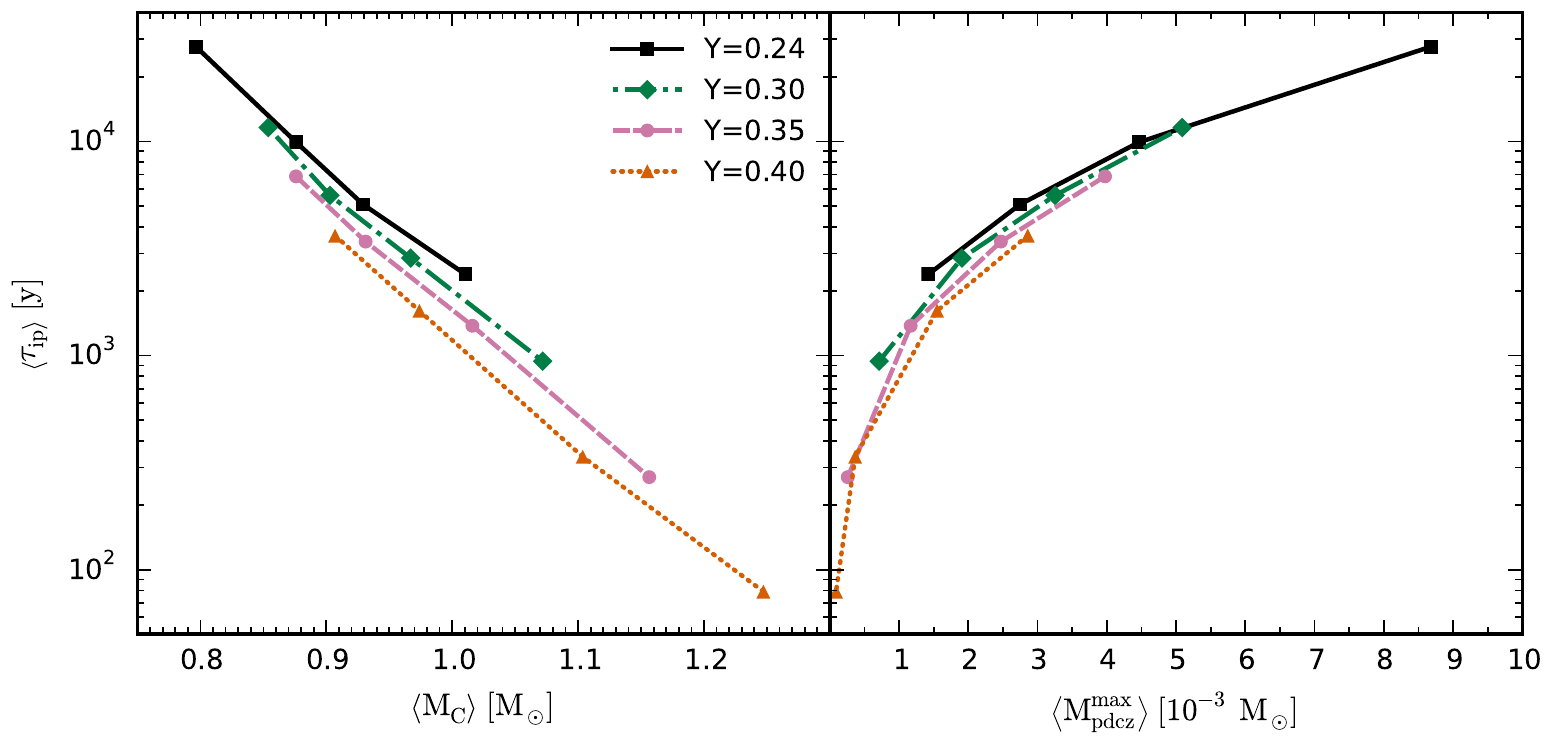}\end{center}
 \caption{Left: The average interpulse period versus the average H-exhausted core mass during the thermally pulsing AGB phase. Right: The mean interpulse period versus the mean of the maximum extent of the pulse-driven convective zone (a proxy for the He-intershell size). Included on the plot are the 3, 4, 5, and 6 \Msun\ models (from left to right) with He mass fractions of $Y= 0.24$ (black squares), 0.30 (green diamonds), 0.35 (pink circles), and 0.40 (orange triangles).}\label{fig:interpulse}
\end{figure*}

To show how the thermal-pulse cycle is altered by He-enhancement, we first measure the average time between thermal pulses (the interpulse). The left panel of Figure \ref{fig:interpulse} shows the average interpulse period versus the average core mass during the AGB for our models from 3 to 6 \Msun\ with He mass fractions of $Y=0.24,0.30,0.35$ and $0.40$. The interpulse time-scale is mainly a function of the H-exhausted core mass \citep{Paczynski:1974kf,ChristySackmann:1975ta}, which increases with the He content \citep{Becker:1979jk,Lattanzio:1986cz}. However, the average interpulse period of a He-enhanced model is lower than that of a primordial-He model with a higher initial mass and the same average core mass. The shorter interpulse time with He-enhancement for a given core mass is connected with the increased H-burning rate. With $Y=0.40$, the average interpulse period is up to 50 per cent shorter than with $Y=0.24$ and the same average core mass. The right panel of Figure \ref{fig:interpulse} shows that a similar, but marginally tighter relation exists when the average core mass is replaced by the average He-intershell mass (approximated by the maximum extent of the pulse-driven convective zone).

\begin{figure}
 \begin{center}\includegraphics[width=\columnwidth]{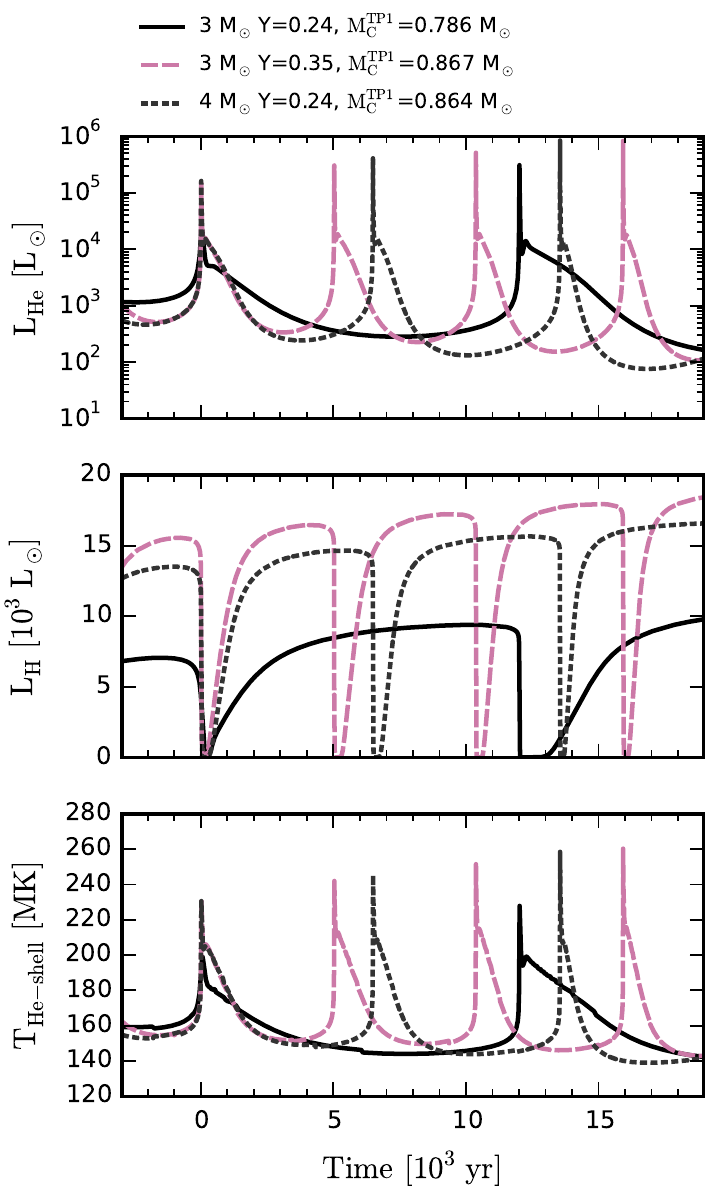}\end{center}
 \caption{He- and H-burning luminosities, and the temperature in the He-burning shell as a function of time after the first thermal pulse for 3 \Msun\ models at $Y=0.24$ and $0.35$, and the 4 \Msun, $Y=0.24$ model.}\label{fig:tp1}
\end{figure}

Figure \ref{fig:tp1} shows the H-burning luminosity and the temperature and luminosity of the He-burning shell as a function of time from just prior to the first thermal pulse until after the second thermal pulse (of the $Y=0.24$ case). To illustrate how a He-rich initial composition alters the behaviour of the burning shells and reduces the interpulse time-scale, the models shown are the 3 \Msun\ model with $Y=0.24$, the 3 \Msun\ model with $Y=0.35$, and the 4 \Msun\ model with $Y=0.24$. These three models enable us to separately compare the effect of increasing the He mass fraction and increasing the initial mass, which each individually lead to a larger core mass. The 3 \Msun, $Y=0.35$ model and the 4 \Msun, $Y=0.24$ model, which have very similar core masses ($0.865 \pm 0.002\ \Msun$) and intershell masses ($5\times10^{-3}\ \Msun$) at the time of their first thermal pulses also exhibit a very similar time evolution of their He-shell temperature and luminosity. However, even with near-identical core masses, the onset of the next thermal pulse takes place sooner in the He-enhanced model, which has a higher H-burning luminosity.

The results in Tables \ref{tbl:structure} and \ref{tbl:structureagb} are derived from our evolutionary sequences which end due to convergence difficulties before the H envelopes are completely ejected \citep{Sweigart:1999tt,Lau:2012ef}. The AGB lifetimes have been calculated from the time spent on the AGB up to the end of our calculations, plus an estimate of the time left to lose the remaining H envelope. This is estimated by dividing the envelope mass at the last computed model by the average mass-loss rate over the last few thousand models. For the number of thermal pulses neglected by our calculations due to the early termination of the AGB, an upper limit is obtained by dividing the envelope ejection time by the average interpulse time. The number of additional thermal pulses is less than 15 per cent of the numbers given in Table \ref{tbl:structureagb} for all cases except for the 6 \Msun, $Y=0.40$ model, which could experience up to $\sim$ 35 per cent more thermal pulses beyond the end of our calculations.

The total number of thermal pulses experienced by each model depends on both the interpulse period and the time spent in the AGB phase, which is controlled by the mass-loss rate. The number of thermal pulses has implications for the stellar yields, as a smaller number of thermal pulses means that there will be fewer neutron-producing events and typically fewer TDU episodes. For models with hot bottom burning \citep{Iben:1975js,Renzini:1981ul,Boothroyd:1992gi,Lattanzio:1992uz}, the length of time spent in the AGB phase also has an impact on the surface abundances and yields of elements involved in proton-capture reactions at the base of the envelope. In the He-rich AGB models, the mass-loss rates are higher due to higher luminosities and generally longer pulsation periods, and thus they have shorter AGB lifetimes. Our models with $Y=0.40$ have shorter AGB lifetimes compared with $Y=0.24$ by a factor of four at 3 \Msun, and by a factor of ten at 6 \Msun, although these results are dependent on the mass loss prescription.

The models presented here for $Y=0.24$ experience significantly more thermal pulses than the models of SCP14 with the same initial mass. For 3, 4, 5, and 6 \Msun, we find 26, 82, 121, and 136 thermal pulses, compared with 15, 23, 35, and 72 for SCP14, and 75, 91, 103, and 116 for VD09. The smaller number of thermal pulses in the models of SCP14 is largely due to their modification of the \citet{Vassiliadis:1993jk} mass-loss prescription to include a minimum rate of $10^{-7.7}$ \Msun/yr at log P $< 2.7$ \citep[described in][]{Straniero:2006do}. VD09 use the \citet{Bloecker:1995ui} mass-loss prescription (with $\eta_R=0.02$), which steeply increases with luminosity. For this reason, while the 3 and 4 \Msun\ models of VD09 experience a greater number of thermal pulses than our models, the opposite is the case with 5 \Msun\ and higher masses, for which our models experience a greater number of thermal pulses.

For the 3 \Msun\ models, variation to the initial He abundance has a small (less than ten per cent) effect on the total number of thermal pulses for $Y=0.24,0.30,$ and $0.35$. However, with $Y=0.40$, the number of thermal pulses almost doubles to 48 in comparison with 26 for $Y=0.35$. The large increase in the number of thermal pulses between $Y=0.35$ and $Y=0.40$ is due to the AGB lifetime shortening by about 15 per cent, while the average interpulse time drops by roughly 50 per cent.

We use the standard dredge-up efficiency parameter defined by $\lambda = \Delta\M{dredge}/\Delta\Mc$, where $\Delta\M{dredge}$ is the mass of He-intershell material mixed to the surface by TDU and $\Delta\Mc$ is the growth of the H-exhausted core during the preceding interpulse phase.

\begin{figure}
 \begin{center}\includegraphics[width=\columnwidth]{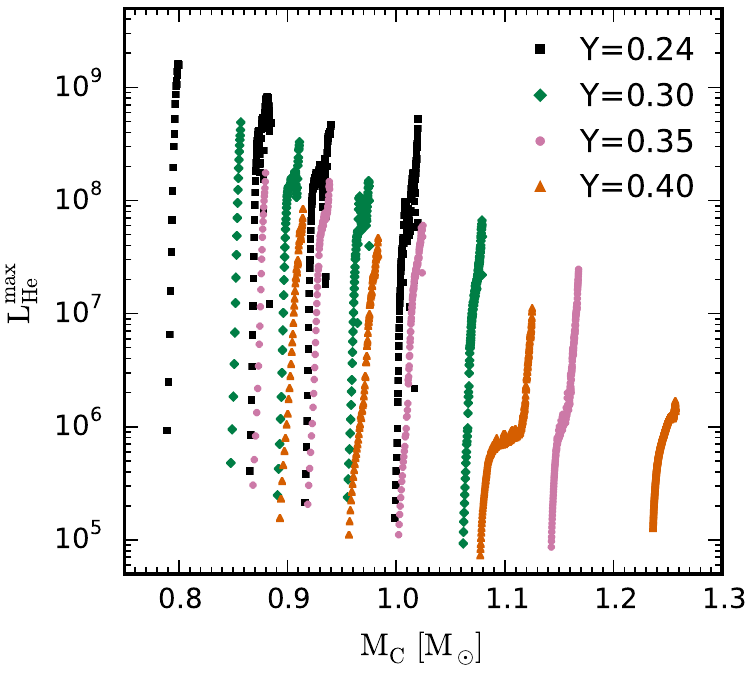}\end{center}
 \caption{The maximum He-burning luminosity versus the H-exhaused core mass for each thermal pulse of the models at 3, 4, 5, and 6 \Msun\ (from left to right) with He mass fractions of $Y= 0.24$ (black squares), 0.30 (green diamonds), 0.35 (pink circles), and 0.40 (orange triangles).}\label{fig:coremass-lumhe}
\end{figure}

The TDU is generally less efficient after weaker thermal pulses, and models with a higher core mass attain a lower peak He-flash luminosity \citep{Sackmann:1980dn,Boothroyd:1988jl,Straniero:2003ho}. This is true for changes to either the initial mass or the initial He mass fraction (Figure \ref{fig:coremass-lumhe}). However, when models of the same core mass are compared, the peak He-burning luminosity in the He-enhanced models is less than the primordial-He models by up to about 80 per cent. As a consequence, the TDU is less efficient in He-enhanced models.

\begin{figure}
 \begin{center}\includegraphics[width=\columnwidth]{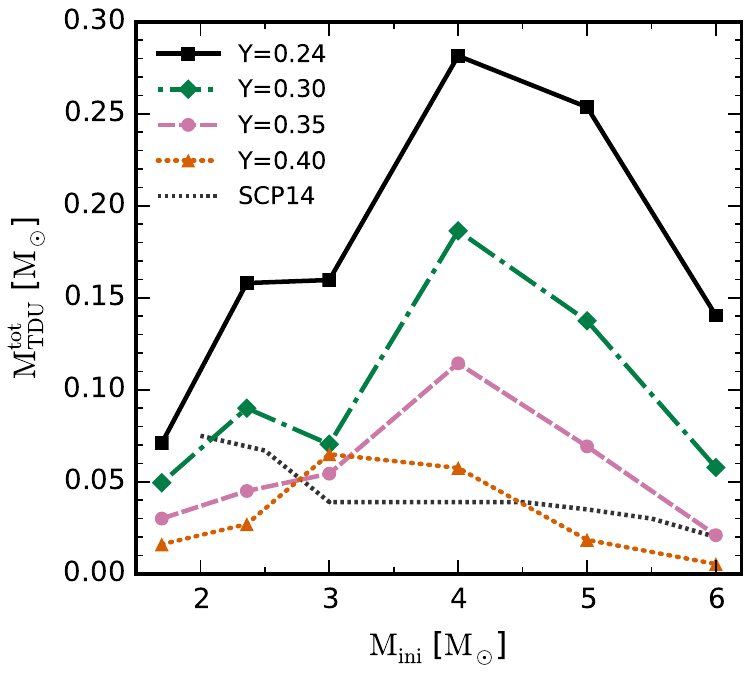}\end{center}
 \caption{The total third dredge-up mass during the AGB as a function of initial mass with He mass fractions of $Y= 0.24$ (black squares), 0.30 (green diamonds), 0.35 (pink circles), and 0.40 (orange triangles). Results of the 1.7 \Msun\ and 2.36 \Msun\ models are from KMN14.}\label{fig:tdumass}
\end{figure}

Figure \ref{fig:tdumass} shows the total mass dredged-up by TDU (hereafter TDU mass; equal to the sum of $\Delta\M{dredge}$) during the AGB phase as a function of initial mass for several initial He abundances. Increasing the initial He mass fraction typically reduces the TDU mass for initial masses from 3 to 6 \Msun, similar to what KMN14 found with their He-enhanced 1.7 \Msun\ and 2.36 \Msun\ models. For $Y=0.24,0.30$, and $0.35$, the maximum dredge-up mass is obtained with an initial mass of around 4 \Msun. With $Y=0.40$ however, the distribution changes shape with a peak shifting downwards to approximately 3 \Msun. In contrast with our results, the SCP14 models predict that the TDU mass decreases monotonically with initial mass, and is overall much lower than for our $Y=0.24$ models (by a factor of up to seven at 4 \Msun). The large variations between results of SCP14 and the present work reflect the significant modelling uncertainties that affect the efficiency of the third dredge-up \citep[for a comparison of TDU efficiency between different evolution codes, see][]{Lugaro:2003ew}. Our specific predictions for the TDU efficiency are influenced by the mass and time resolution of the extremely He-enhanced models, and our treatment of the border between radiative and convective zones \citep[as demonstrated by][]{Frost:1996fu}.

The TDU mass of the 3 \Msun, $Y=0.40$ model is unusual because it exceeds the TDU mass of the 3 \Msun\ model with $Y=0.35$ despite the increased He abundance, which reduces the dredge-up efficiency. Indeed, both the average and maximum values of $\lambda$ are lower in the 3 \Msun\ model with $Y=0.40$ than $Y=0.35$ (Table \ref{tbl:structureagb}). The unusually high TDU mass with $Y=0.40$ is the result of a near-doubling in the number of thermal pulses, which leads to more thermal pulses with TDU.

\subsection{Hot bottom burning}\label{sec:hbb}
At $Z=0.0006$ and primordial He abundance, AGB stars with initial masses $\gtrsim 4$ \Msun\ experience sufficiently high temperatures at the base of the convective envelope (\Tsub{bce}) for proton-capture nucleosynthesis to take place there. This is called hot bottom burning (HBB).

With temperatures in the envelope above about 50 MK, the resulting activation of the CN cycle begins to convert a significant fraction of C nuclei into N. In the absence of other effects, the CN-cycle would cause the C/O ratio in the envelope to decrease. However, additional primary \iso{12}{C} nuclei from the He-burning shell are periodically transported to the envelope via TDU, leading to a surface C/O ratio that depends on the interplay between TDU and HBB. In the more massive AGB stars at solar metallicity, HBB is observationally confirmed to prevent the surface abundances from becoming C-rich \citep[C/O $>$ 1;][]{Lattanzio:1992uz,Boothroyd:1993dj}. Further, the combined operation of both TDU and HBB in nature is confirmed by observations of C-deficient, N-rich AGB stars in the Magellanic clouds \citep{McSaveney:2007ks}.

\begin{figure}
 \begin{center}\includegraphics[width=\columnwidth]{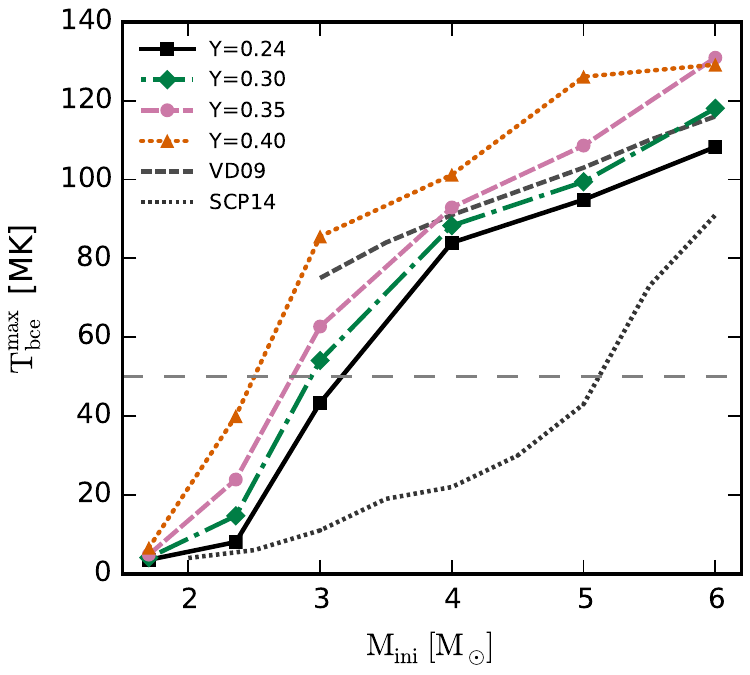}\end{center}
 \caption{The maximum temperature at the base of the convective envelope during the interpulse phase as a function of initial mass with He mass fractions of $Y= 0.24$ (black squares), 0.30 (green diamonds), 0.35 (pink circles), and 0.40 (orange triangles). Results of the 1.7 \Msun\ and 2.36 \Msun\ models are from KMN14. The horizontal dashed line at 50 MK indicates the approximate temperature above which hot bottom burning significantly alters surface abundances.}\label{fig:maxtbce}
\end{figure}

Figure \ref{fig:maxtbce} shows the maximum temperature at the base of the convective envelope in our models as a function of the initial mass. The higher temperatures in models with increased initial He abundance enables HBB to take place at lower initial masses than at primordial He abundance. At masses above about 4 \Msun, HBB will proceed at increased rates with He-enhancement, while models of less than about 2 \Msun\ will not experience HBB temperatures even with He abundances as high as $Y=0.40$. For the 3 \Msun\ models, \Ttwo{bce}{max} increases from 43 MK with $Y=0.24$ to 86 MK with $Y=0.40$. This corresponds to a change from virtually no HBB to significant CN cycling in the envelope.

The maximum envelope temperatures of the primordial-He models are in reasonable agreement (less than ten per cent difference) with the models of VD09 for masses of 4 \Msun\ and above. The difference is more significant at 3 \Msun, where our model has a maximum temperature of 57 MK during TDU and 43 MK during the interpulse phase, while the corresponding model of VD09 has a maximum temperature of 75 MK. The higher envelope temperatures in the models of VD09 are a result of the FST convection model, which predicts more efficient convective transport than the MLT theory. In contrast to the models of VD09 and the models presented here, the models of SCP14 predict much lower temperatures at the base of the convective envelope (22 MK at 4 \Msun\ versus 84 MK in our model and 91 MK for VD09). SCP14 compare their models to the models of \citet{DOrazi:2013bq} (calculated with the same evolutionary code used here), which experience more efficient HBB. They attribute the difference to some combination of the equation of state, interpolation of the radiative opacity tables, and their particular mixing scheme.

\begin{figure}
 \begin{center}\includegraphics[width=\columnwidth]{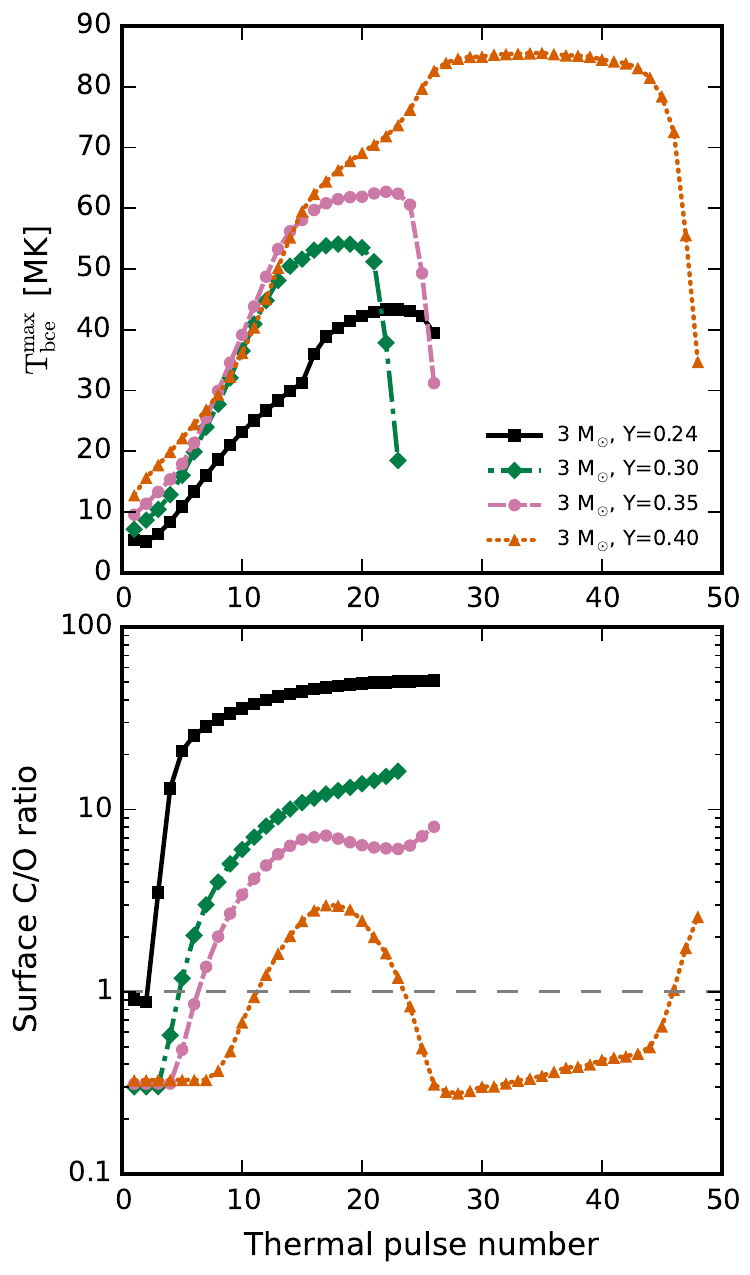}\end{center}
 \caption{The maximum temperature at the base of the convective envelope during the preceding interpulse phase (top panel) and the surface C/O ratio (bottom panel) as a function of thermal pulse number for 3 \Msun\ models with He mass fractions of $Y= 0.24$ (black squares), 0.30 (green diamonds), 0.35 (pink circles), and 0.40 (orange triangles). The dashed line in the lower panel indicates a C/O ratio of 1.}\label{fig:hbb-3msun}
\end{figure}

Figure \ref{fig:hbb-3msun} shows the maximum temperature at the base of the convective envelope and the surface C/O ratio for each thermal pulse of the 3 \Msun\ models. Over the first ten or so TPs, the shallower initial rise of C/O in the He-enhanced models can be attributed to the less efficient TDU, as none of the models have reached HBB temperatures at this stage. In models with He-enhancement, HBB becomes active and is particularly noticeable with $Y=0.35$ and $Y=0.40$, for which significant CNO cycling in the envelope causes the C/O ratio to begin decreasing at around TP 16. As mass loss erodes the envelope to below a critical mass value (around 1.5 \Msun\ in this case) over the last few thermal pulses, HBB ceases while dredge-up continues to take place, causing an upturn in the C/O ratio in these two models \citep[see][]{Frost:1998vn}. In the $Y=0.40$ model, the cessation of HBB causes the surface to transition from O-rich to C-rich (for a second time) at the third-last thermal pulse.

Aside from CNO abundances, HBB can alter the abundances of other light elements through the NeNa cycle and the MgAl chain. Activation of the NeNa cycle can produce or destroy \iso{23}{Na} at the expense of Ne isotopes \citep{Arnould:1999ue,Mowlavi:1999ut,Karakas:2003el}. At temperatures of 50-200 MK, proton captures on to \iso{23}{Na} produce more \iso{20}{Ne} than \iso{24}{Mg}, as the rate of the $\iso{23}{Na}(p,\alpha)\iso{20}{Ne}$ reaction is several times faster than $\iso{23}{Na}(p,\gamma)\iso{24}{Mg}$ \citep{Hale:2004fa}. Activation of the MgAl chain can result in a net production of \iso{26}{Al} and \iso{27}{Al} \citep{Arnould:1999ue,Denissenkov:2003bn,Ventura:2008iv}, and further proton capture in the most massive AGB stars can produce \iso{28}{Si} \citep{Ventura:2011kn}.

\subsection{Core mass-luminosity relation}
\begin{figure}
 \begin{center}\includegraphics[width=\columnwidth]{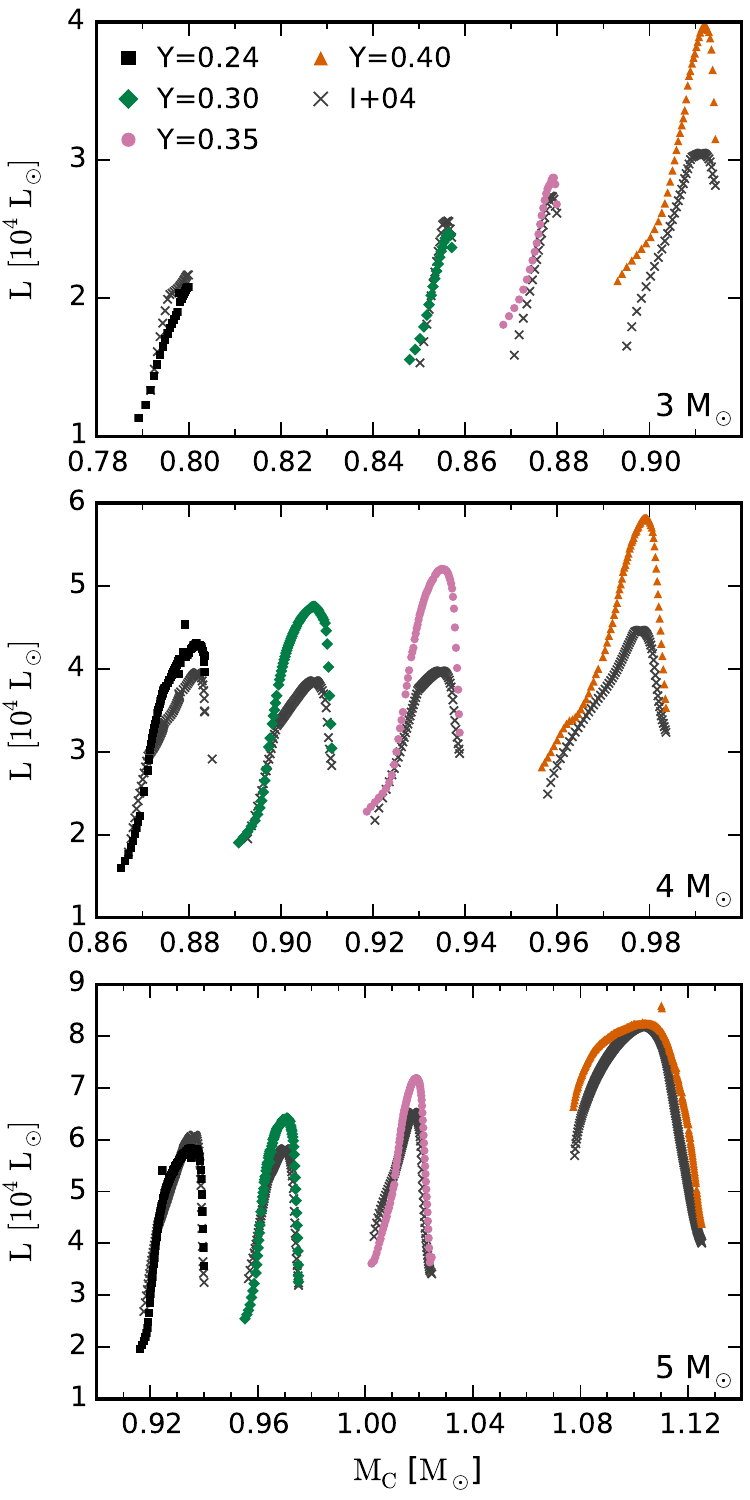}\end{center}
 \caption{Luminosity versus H-exhausted core mass during the AGB for 3, 4, and 5 \Msun\ models with He mass fractions of $Y= 0.24$ (black squares), 0.30 (green diamonds), 0.35 (pink circles), and 0.40 (orange triangles). The plot also shows values from the fitting formula of \citet{Izzard:2004bq} (I+04).}\label{fig:cmlr}
\end{figure}

A well-known correlation exists between the surface luminosity of an AGB star and its H-exhausted core mass. A linear form was first proposed by \citet{Paczynski:1970uk}, although stars undergoing HBB and TDU were found to diverge from this relation \citep{Bloecker:1991tt,Lattanzio:1992uz,Boothroyd:1992gi,Marigo:1999va}. Figure \ref{fig:cmlr} shows the surface luminosity versus core-mass at each thermal pulse for the models of 3, 4, and 5 \Msun.

We also compare the core-mass-luminosity behaviour of our models with values from the fitting formula specified by equations 29 to 34 of \citet{Izzard:2004bq}. This formula estimates the surface luminosity as a function of the core mass, the envelope mass, and the total growth in the core mass during the TP-AGB (neglecting decreases from TDU). The formula is based on the stellar evolutionary models of \citet{Karakas:2002ci} with He mass fractions that range from 0.24 and 0.30, depending on the metallicity. The grid of evolutionary models used to generate the fit extends down to $Z=0.0001$, but only with masses of 1.25 to 2.25 \Msun. At $Z=0.004$ and higher metallicities, the grid includes models with initial masses up to 6 \Msun. Although the updated formula of \citet{Izzard:2006ij} improves the fit to models of low metallicity, its parameterisation in terms of the initial mass causes it to be less accurate for He-rich models, which have different stellar structures compared to primordial-He models at a given initial mass.

For the 3 \Msun\ models, the formula is a good fit for He mass fractions between $Y=0.24$ and extrapolation up to $Y=0.35$. However, our 3 \Msun\ model with $Y=0.40$ is consistently more luminous (by up to 25 per cent) than the formula value. Between $Y=0.35$ and $Y=0.40$, the maximum temperature at the base of the convective envelope during H-burning increases from 63 to 86 MK, and the more extreme HBB further increases the surface luminosity. The excess envelope luminosity with $Y=0.40$ in comparison to the fitting formula (which accounts for the larger core mass) is probably partly due to the more He-rich envelope composition (which affects the H-burning rate) and also because of the smaller envelope mass compared to higher-mass models with primordial He and the same core mass.

The 4 \Msun\ models at all He mass abundances have higher peak luminosities than the formula values, although this is a small effect with $Y=0.24$. The He-enhanced 4 \Msun\ models diverge from the relation in a manner characteristic of HBB, with a rise in the middle followed by a decline as the decreasing envelope mass causes a reduction in the temperature at the base of the envelope. A reasonably good fit is found for the 5 \Msun\ models, and this is probably because HBB is already active for this initial mass with $Y=0.24$.

In summary, the initial He abundance alters the core-mass-luminosity relation, and a more accurate fitting formula for application to He-rich populations would require a fit to stellar evolutionary models at the appropriate He abundance.

\subsection{Carbon burning and final fates}\label{sec:remnants}
The C-burning behaviour and the stellar remnants of the models depend on the initial stellar mass and the initial He abundance. There is also a dependence on the initial metallicity \citep{Cassisi:1993is,Umeda:1999ed}.

All of the models presented in this work would eventually form white dwarfs (WD) rather than exploding as electron-capture supernovae because they end with core masses well below the limiting mass of $1.37\ \Msun$ \citep{Miyaji:1980uz}. In Table \ref{tbl:structure} we classify the stellar remnants for each of our models based on the core composition at the end of our calculations, which would be similar to the composition of the resulting WD remnant.

AGB stars with initial masses below the minimum mass required for off-centre core C ignition \citep[\M{up}, as defined by][]{Becker:1979jk} end the AGB with cores composed largely of C and O in roughly equal proportions that are set by earlier core and shell He burning. Thus, we classify the remnants from our models without C ignition as CO WDs.

In stars with degenerate CO cores following central He exhaustion, the position of maximum temperature moves outwards from the centre to where the energy liberated by gravitational contraction exceeds the rate of cooling by neutrino emission \citep{Becker:1979jk}. In stars more massive than \M{up}, the temperature maximum exceeds about 600 to 700 MK \citep{Siess:2006ki} and C is ignited off-centre in the core. In this case, the star is referred to as a `super-AGB' star \citep{GarciaBerro:1994kp,Siess:2007ih,Siess:2010hj}.

C burning proceeds via the $\iso{12}{C}(\iso{12}{C}, \alpha)\iso{20}{Ne}$ and $\iso{12}{C}(\iso{12}{C}, p)\iso{23}{Na}$ reactions. The latter reaction is followed by $\iso{23}{Na}(p,\alpha)\iso{20}{Ne}$, and thus also contributes to the abundance of \iso{20}{Ne} in the core, as does the $\iso{16}{O}(\alpha,\gamma)\iso{20}{Ne}$ reaction. The region affected by C burning will be composed largely of O and Ne, with a ratio that is affected by the rate of $\iso{16}{O}(\alpha,\gamma)\iso{20}{Ne}$. The C burning can either be aborted before reaching the centre and form a hybrid CO(Ne) WD \citep{Doherty:2010fk,Karakas:2012kc,Denissenkov:2013dd}, or proceed all the way to the centre and form an ONe WD.

\citet{Bono:2000gk} show for super metal-rich compositions ($Z=0.04$) that the $\M{up}$ boundary decreases with increasing He abundance from $9.5 \pm 0.5\ \Msun$ for $Y=0.29$ to $7.7 \pm 0.2\ \Msun$ for $Y=0.37$. With primordial He abundance, C burning at $Z=0.0006$ takes place with a minimum initial mass of 6.5 \Msun\ \citep{Doherty:2015ek}, and we do not find C burning in our $Y=0.24$ models with masses up to 6 \Msun. These models lead to CO WDs. The 6 \Msun\ model with $Y=0.30$ experiences off-centre C burning that leaves C in the inner 0.5 \Msun\ unburned, and this region remains composed of C and O. We classify the remnant of this model as a hybrid CO(Ne) white dwarf \citep{Doherty:2015ek}. The models at 5 \Msun\ and below with $Y=0.30$ do not show any significant C burning.

\begin{figure}
 \begin{center}\includegraphics[width=\columnwidth]{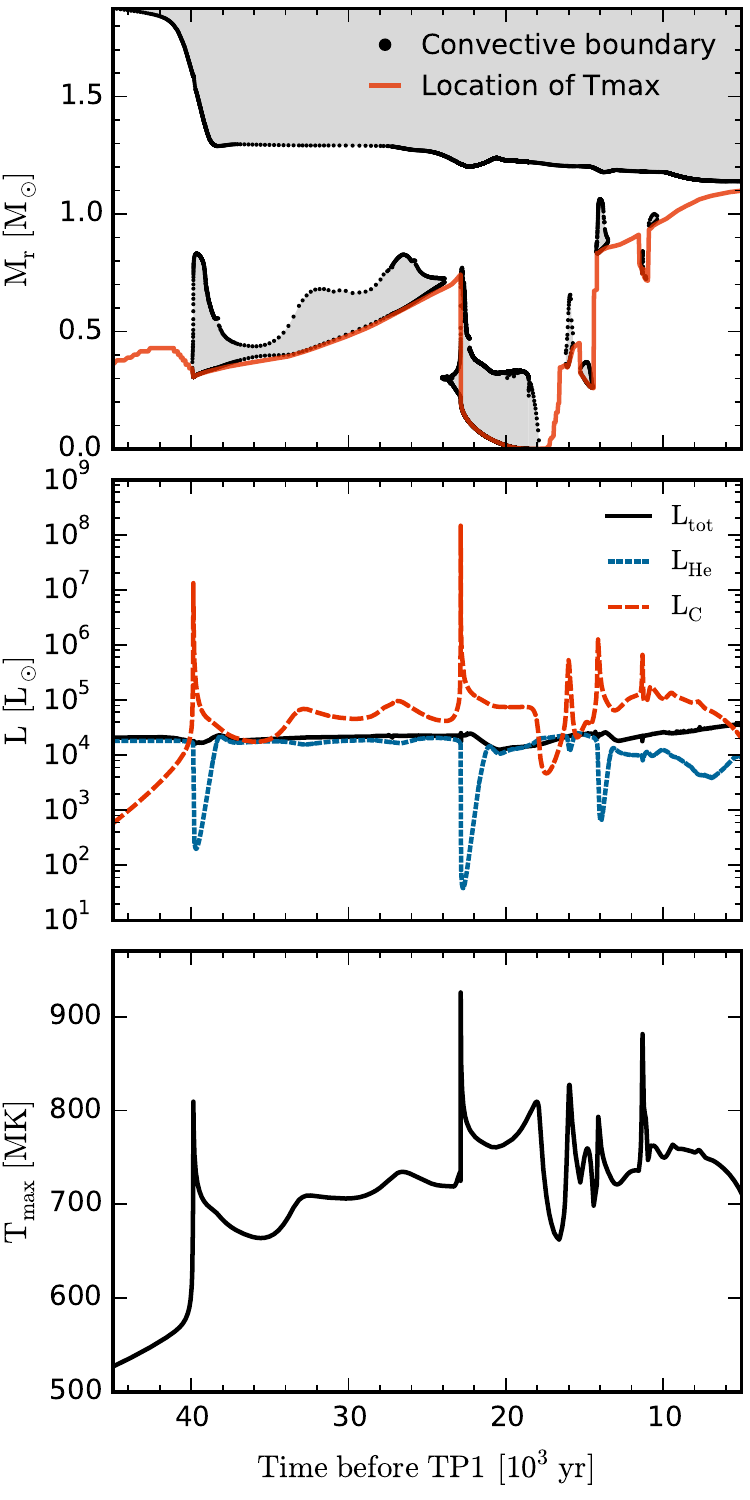}\end{center}
 \caption{C-burning characteristics of the 6 \Msun, $Y=0.35$ model versus time before the first thermal pulse. Top: A Kippenhahn diagram showing C-burning and envelope convective zones and the location of maximum temperature (red). Middle: The total luminosity ($\mathrm{L}_\mathrm{tot}$) and the luminosities due to He- and C-burning ($\Lsub{He}$ and $\Lsub{C}$). Bottom: The maximum temperature in the model.}\label{fig:cburning}
\end{figure}

For $Y=0.35$, the 6 \Msun\ model experiences off-centre C ignition that proceeds to full central C burning and forms an ONe core. Figure \ref{fig:cburning} shows the C-burning luminosity and the behaviour of the C-burning convective zones in this model as a function of time. For $Y=0.35$, the models at 5 \Msun\ and below do not show any significant C burning. For $Y=0.40$, the 5 \Msun\ model experiences off-centre C burning that produces a CO(Ne) white dwarf and the 6 \Msun\ model undergoes full central C burning that produces an ONe white dwarf. The models with $Y=0.40$ at 4 \Msun\ and below do not show any significant C burning.

In summary, we find that \M{up} at $Z=0.0006$ exceeds 6 \Msun\ for $Y=0.24$, decreases to between 5 and 6 \Msun\ for $Y=0.30$ to $0.35$, and decreases further to between 4 and 5 \Msun\ with $Y=0.40$. Future investigations with a finer mass grid would be required to determine more precisely the dependence of \M{up} on the He abundance.

\section{Nucleosynthesis and Stellar Yields}\label{sec:nucleosynthesisyields}
The stellar yield represents the contribution of a particular chemical species $i$ (e.g., an element or nuclide) to the interstellar medium by a star over its lifetime, and is calculated using the formula
\begin{equation}
 M_{i}^\mathrm{yield} = \int_{0}^{\tau} X_i(t) \dot{M}(t) \,\mathrm{d}t,
\end{equation}
where $X_i(t)$ is the mass fraction of species $i$ at time $t$ and $\dot{M}(t)$ is the stellar mass-loss rate at time $t$. For models with a non-zero envelope mass at the end of our calculations, we assume that all mass exterior to the core is ejected with the composition of the surface at the last computed model.

The stellar yields of all models are provided as online data tables, with an example of their form and content in Table \ref{atbl:elementalyields}. In the online data tables, we also include the net yields, which have the initial abundances subtracted according to
\begin{equation}
 M_{i}^\mathrm{netyield} = \int_{0}^{\tau} \left[ X_i(t) - X_i(0) \right] \dot{M}(t) \,\mathrm{d}t,
\end{equation}
where $X_i(0)$ is the mass fraction of species $i$ in the initial composition. The net yields indicate whether a chemical species is produced or depleted in the context of chemical evolution.

\subsection{Yields of light elements}\label{sec:resultslightelements}
\begin{figure}
 \begin{center}\includegraphics[width=\columnwidth]{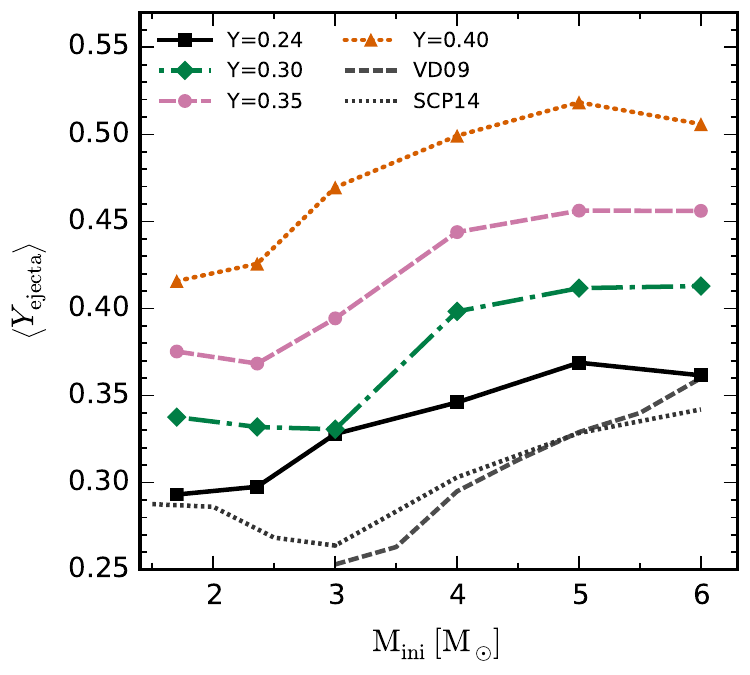}\end{center}
 \caption{Average He mass fractions in the ejecta of the stellar models versus initial mass with He mass fractions of $Y= 0.24$ (black squares), 0.30 (green diamonds), 0.35 (pink circles), and 0.40 (orange triangles). 1.7 \Msun\ and 2.36 \Msun\ models are from KMN14.}\label{fig:yieldhelium}
\end{figure}

In AGB stars, the primary He produced by H-burning reactions is mixed to the surface during dredge-up events, with the largest increase to the surface abundance of He occurring during SDU and a smaller increase due to TDU. As we show in Figure \ref{fig:yieldhelium}, the average He mass fractions in the ejecta of our models are typically 0.05 to 0.10 higher than the initial He mass fractions, with the higher-mass models releasing material that is more He-enhanced due to a deeper SDU \citep{Karakas:2006fg}.

For primordial-He models, the He mass fractions of our yields are generally higher than the predictions of SCP14 or VD09. At the largest difference, our predictions for $\left<Y_\mathrm{ejecta}\right>$ are higher than VD09 and SCP14 by about 0.05 and 0.03, respectively. The FRUITY database \citep{Cristallo:2011fz} provides surface He mass fractions of the SCP14 models after each dredge-up event, which enables us to compare the effects of SDU and TDU. At 3 \Msun, the surface $Y$ following SDU is 0.30 (cf., 0.26, SCP14) and increases with each TDU episode up to 0.33 at the end of our calculations, while the model of SCP14 finishes with a surface $Y$ of 0.27. The higher He mass fraction in our yield at 3 \Msun\ compared with SCP14 is caused by a deeper SDU and more extensive TDU in our model. For initial masses of 4 to 6 \Msun, our post-SDU He abundances are very similar to the models of SCP14, and the differences in $\left<Y_\mathrm{ejecta}\right>$ largely reflect differences in the efficiency of HBB.

\begin{figure*}
 \begin{center}\includegraphics[width=\textwidth]{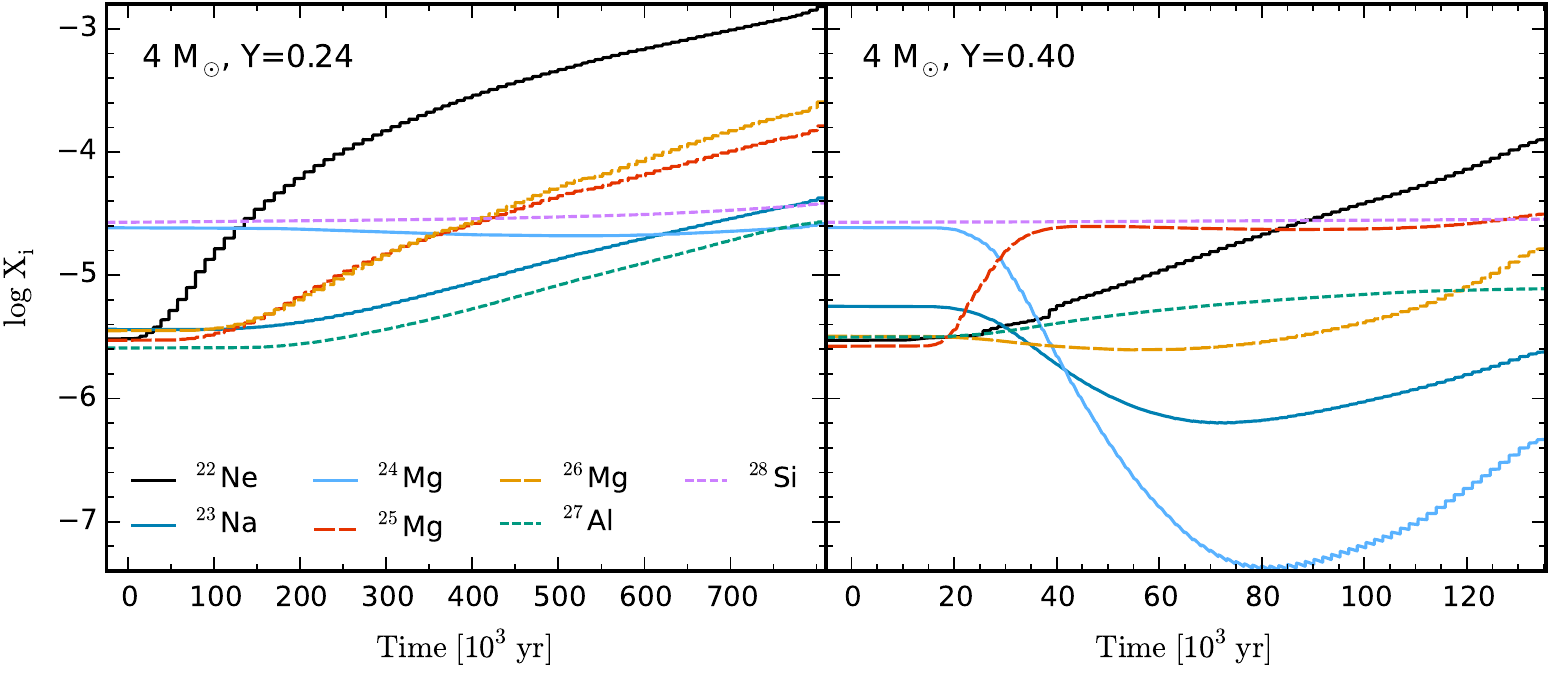}\end{center}
 \caption{Surface abundance of selected nuclides that participate in the NeNa cycle and the MgAl chain versus time in 4 \Msun\ models at $Y=0.24$ (left panel) and $Y=0.40$ (right panel). Time on the horizontal axis is relative to the time of the first thermal pulse.}\label{fig:surflightelm}
\end{figure*}

Figure \ref{fig:surflightelm} illustrates the time evolution of the surface abundances of Ne, Na, Mg, Al, and Si isotopes in the 4 \Msun\ models with $Y=0.24$ (left panel) and $Y=0.40$ (right panel). We focus on the 4 \Msun\ models as they demonstrate the significant changes that can occur due to HBB with variations to the initial He abundance. As shown in Table \ref{tbl:structureagb}, the maximum temperature at the base of the envelope during the interpulse phase increases from 84 MK at $Y=0.24$ to 101 MK at $Y=0.40$, with implications for the nucleosynthesis via proton capture reactions. However, while the maximum temperature increases, other important factors include the length of time during which the envelope is subject to high temperatures and the total mass of TDU ($t_\mathrm{HBB}$ and $\Mtwo{TDU}{tot}$ in Table \ref{tbl:structureagb}), both of which are lower in the $Y=0.40$ model.

The surface abundance of \iso{22}{Ne} displayed in Figure \ref{fig:surflightelm} for both $Y=0.24$ and $Y=0.40$ follows a stair-step increase as \iso{22}{Ne} is periodically dredged up into the envelope following its production in convective pulses. The surface abundance of \iso{23}{Na}, which increases monotonically at $Y=0.24$, is initially destroyed by proton captures in the envelope at $Y=0.40$. Nucleosynthesis in the convective pulses also contributes to the intershell abundances of \iso{25}{Mg} and \iso{26}{Mg}, which are produced via the $\iso{22}{Ne}(\alpha,n)\iso{25}{Mg}$ and $\iso{22}{Ne}(\alpha,\gamma)\iso{26}{Mg}$ reactions. At $Y=0.24$, increases to \iso{25}{Mg} and \iso{26}{Mg} occur during dredge-up events, with minimal increases due to HBB. At $Y=0.40$, however, proton captures cause the \iso{25}{Mg} abundance to increase rapidly at the expense of \iso{24}{Mg}. While some of the \iso{24}{Mg} is destroyed by proton captures at $Y=0.24$ (0.1 dex in mass fraction), the destruction of \iso{24}{Mg} is far more extensive at $Y=0.40$, with a decrease by almost 3.0 dex before the reduced envelope temperatures and efficient dredge-up cause the abundance to begin increasing. The production of \iso{27}{Al} is lower at $Y=0.40$ despite the higher $\Ttwo{bce}{max}$, which indicates the greater impact of a shorter AGB phase and reduced dredge-up in the He-rich model for this nuclide. In both cases, the amount of \iso{28}{Si} produced from proton capture on to \iso{27}{Al} is very small ($< 0.1$ dex).

\begin{figure*}
 \begin{center}\includegraphics[width=\textwidth]{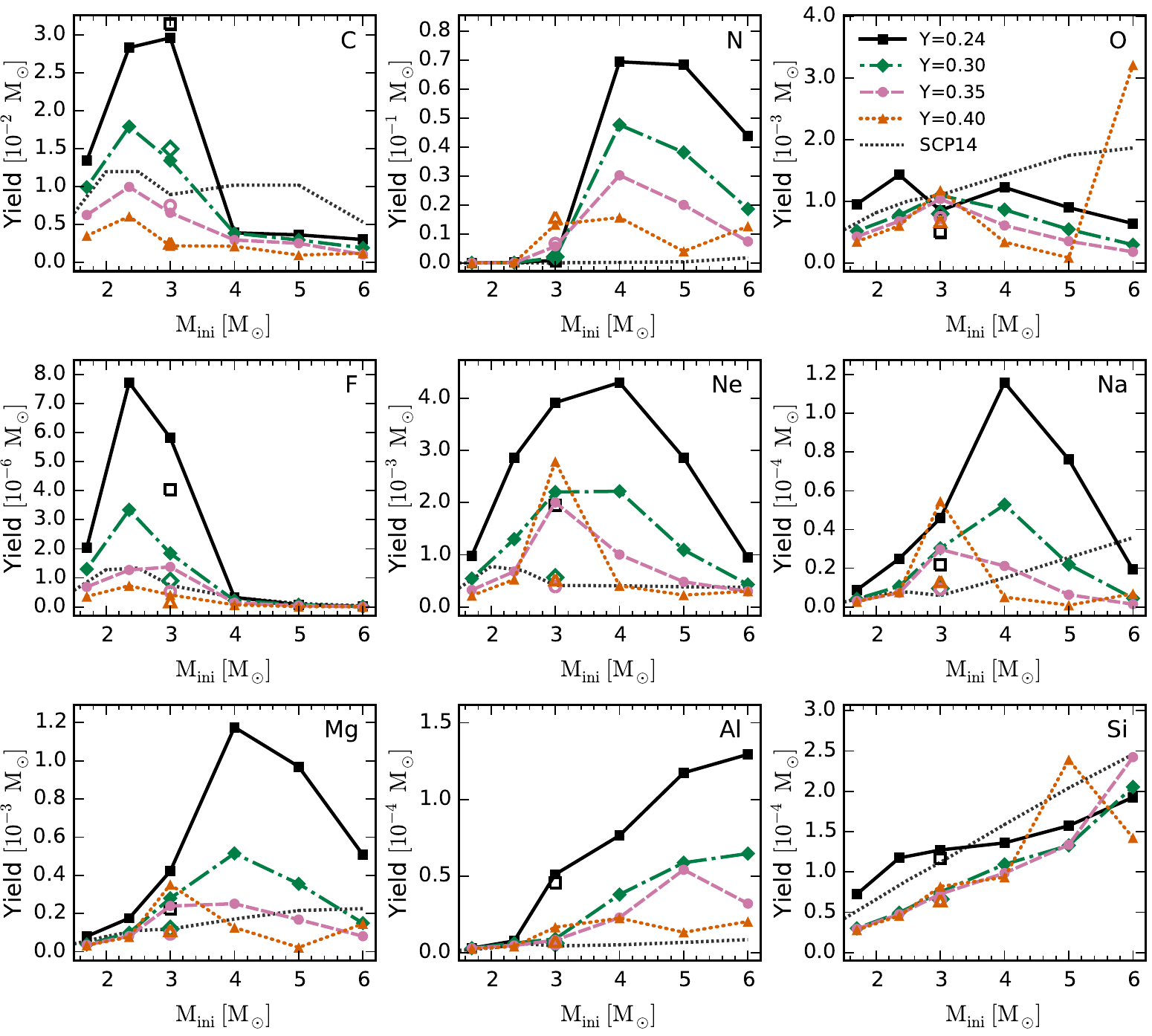}\end{center}
 \caption{Yields of selected light elements as a function of initial mass with He mass fractions of $Y= 0.24$ (black squares), 0.30 (green diamonds), 0.35 (pink circles), and 0.40 (orange triangles). Open points represent 3 \Msun\ models without a PMZ. Models with masses of 3 \Msun\ and below otherwise include a PMZ of $1\times10^{-3}$ \Msun. 1.7 \Msun\ and 2.36 \Msun\ yields are from KMN14.}\label{fig:yieldslightelvsmass}
\end{figure*}

Figure \ref{fig:yieldslightelvsmass} shows the stellar yields (in \Msun) of the light elements C, N, O, F, Ne, Na, Mg, and Al as a function of initial mass.

At primordial He abundance, the yield of C increases with the initial mass (and TDU mass) up to a peak at around 3 \Msun. This is the highest mass model that does not experience HBB, while HBB is highly effective in reducing the C yield and increasing the yield of N with masses of about 4 \Msun\ and above. With helium mass fractions from 0.30 to 0.40, the model that produces the highest C yield becomes the 2.36 \Msun\ model of KMN14. For the models of 3 \Msun, the increase in the N yield with He abundance shows that the increased efficiency of HBB with He-enhancement overcomes the opposing effect of reduced TDU. However, this changes with M $\gtrsim 4$ \Msun, where the yields of N decrease with increasing He, demonstrating that the reduced TDU and shorter AGB phases have a greater effect on the N yields than the increased temperatures at the base of the envelope.

For most of the models, the sum of the C+N+O yields is substantially reduced by helium enhancement, with the exception of the 6 \Msun\ model with $Y=0.40$. The 6 \Msun\ model with $Y=0.40$ experiences a corrosive SDU that dredges up C and O from the core, which affects the stellar yields of several light elements. Compared with the 6 \Msun\ model at $Y=0.35$ (which does not experience a corrosive SDU), the yield of N (produced from dredged-up C) increases by 70 per cent, while O increases by 1600 per cent and the Mg yield increases by 80 per cent.

The predicted yields of C for the models of SCP14 are up to three times lower than our $Y=0.24$ models at masses of 3 \Msun\ and below, and roughly two times higher for masses of 4 \Msun\ and above. The smaller C yields in the lower mass range reflect the significantly reduced TDU masses of the SCP14 models, which are up to 85 per cent lower than the models presented here. The higher C yields of the SCP14 models in the upper mass range occur in spite of the reduced TDU and are caused by lower envelope temperatures and consequently less HBB than the models presented here.

The only stable isotope of F is \iso{19}{F}, which is produced via $\iso{14}{N}(\alpha,\gamma)\iso{18}{F}(\beta^+\nu)\iso{18}{O}(p, \alpha)\iso{15}{N}(\alpha, \gamma)\iso{19}{F}$ during thermal pulses in the He-intershell \citep{Jorissen:1992us}. F can be destroyed both by proton captures in the envelopes of stars with HBB via $\iso{19}{F}(p,\alpha)\iso{16}{O}$ and $\alpha$ captures in the He-intershells of stars more massive than about 5 \Msun\ via $\iso{19}{F}(\alpha,p)\iso{22}{Ne}$ \citep{Lugaro:2004en,Cristallo:2014ev}. Thus, the F yields depend on both the TDU mass and the temperatures at the base of the convective envelope, causing a narrow peak of production at around 2 to 3 \Msun\ that makes F abundances a powerful indicator of stellar mass \citep[e.g.,][]{DOrazi:2013dm}.

The yield of Ne is sensitive to the inclusion and extent of a PMZ \citep{Shingles:2013kg}. This is because additional primary \iso{14}{N} produced in the PMZ is converted into Ne during convective pulses via \iso{14}{N}$(\alpha,\gamma)$\iso{18}{F}$(\beta^+\nu)$\iso{18}{O}$(\alpha,\gamma)$\iso{22}{Ne}. The very short half-life of \iso{18}{F} (110 min) means that the additional F nuclei produced through this chain will not reach the surface (unless neutrons are captured to make \iso{19}{F}). The increase in the yield of F with the inclusion of a PMZ is caused by the additional production of \iso{15}{N}, which is converted into \iso{19}{F} during thermal pulses.

\begin{figure}
 \begin{center}\includegraphics[width=\columnwidth]{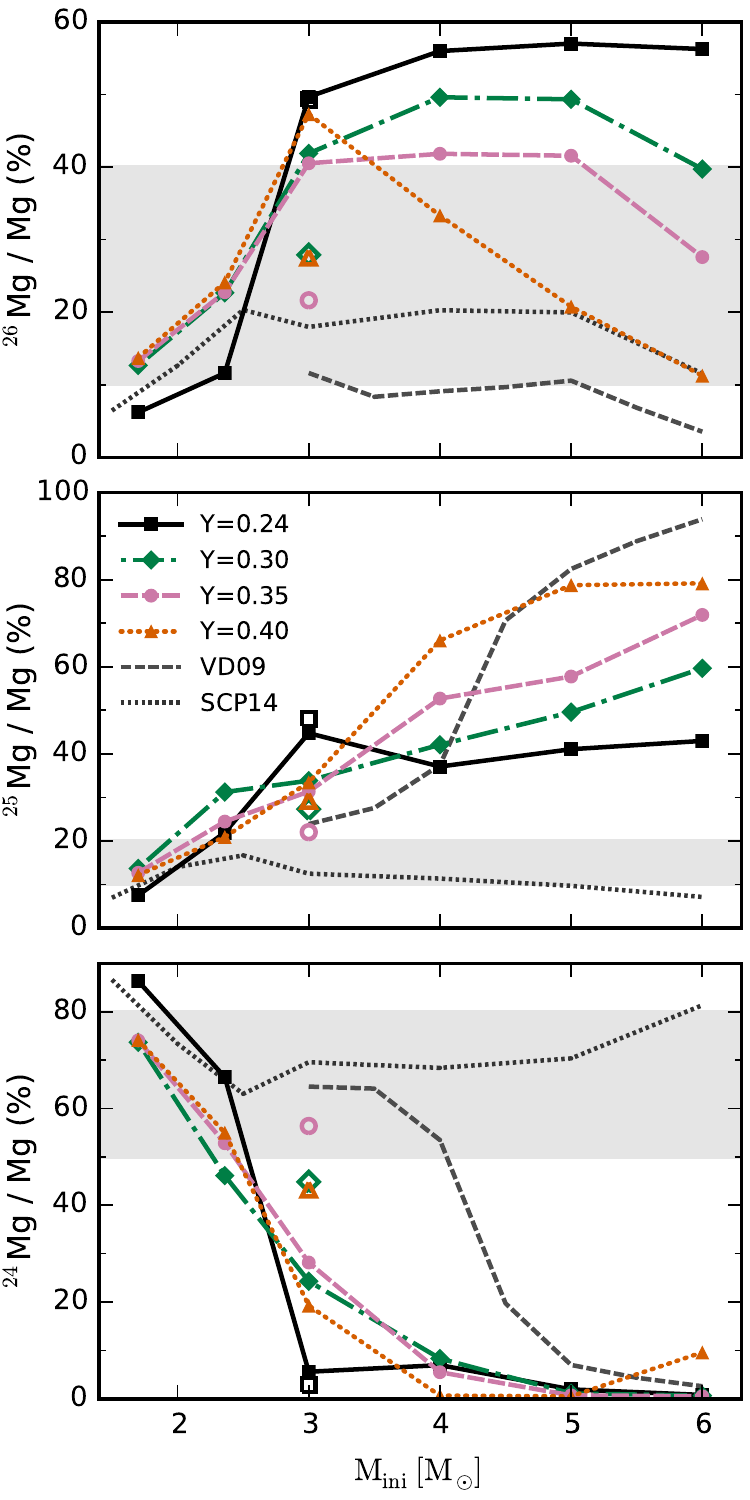}\end{center}
 \caption{The isotopic fractions (by number) of Mg in the stellar yields as a function of initial mass for He mass fractions of $Y= 0.24$ (black squares), 0.30 (green diamonds), 0.35 (pink circles), and 0.40 (orange triangles). Open points indicate 3 \Msun\ models without a PMZ. Models with initial masses $\leq 3$ \Msun\ otherwise include a PMZ of $1\times10^{-3}$ \Msun. The points at 1.7 \Msun\ and 2.36 \Msun\ are from the models of KMN14 (which have [$\alpha$/Fe] $=0.4$ for $Y=0.24$). The grey shaded regions indicate the approximate range of red giants in $\omega$ Centauri with [Fe/H] $\gtrsim -1.4$ from observations by \citet{DaCosta:2013cf}.}\label{fig:mgisotopes}
\end{figure}

Figure \ref{fig:mgisotopes} shows the isotopic fractions of \iso{24}{Mg}, \iso{25}{Mg}, and \iso{26}{Mg} in the stellar yields as a function of the initial stellar mass and He abundance. The shaded regions indicate the approximate range of Mg isotopic fractions for red giants in $\omega$ Centauri with [Fe/H] $\gtrsim -1.4$, as measured by \citet{DaCosta:2013cf}.

\iso{24}{Mg} is efficiently produced by the $\alpha$-process in core-collapse supernovae, while the presence of \iso{25}{Mg} and \iso{26}{Mg} at low metallicities is generally a good indicator of enrichment by massive AGB stars or Wolf--Rayet stars. In the $\omega$ Centauri stars, the range of \iso{24}{Mg}/Mg fractions of about 50 to 80 per cent is lower than would be expected from a pure core-collapse enrichment scenario  \citep[e.g., $\gtrsim 90$ per cent for the low-metallicity yields of][]{Kobayashi:2011hj} and includes values well below the solar value of 79 per cent \citep{Asplund:2009eu}. However, a precise accounting for the contribution by AGB stars to these low \iso{24}{Mg}/Mg values in $\omega$ Centauri would require a detailed chemical evolution model \citep[e.g.,][]{Romano:2007ks}, and it is not immediately clear whether the yields of our He-rich models provide a better match to these observations.

This use of our yields or those of VD09 could potentially constrain the mass range of AGB polluters in $\omega$ Centauri. This is because the Mg isotope ratios in these yields vary strongly as a function of the initial stellar mass. In contrast, the limited HBB in the models of SCP14 leads to low isotopic fractions of \iso{25}{Mg} and \iso{26}{Mg}, which remain similar to the values in the initial composition throughout the mass range from 1 to 6 \Msun. These models predict that Mg isotope ratios provide only a very weak constraint on the mass range of AGB polluters.

In our models for $Y=0.40$, the \iso{26}{Mg}/Mg ratio sharply decreases with initial masses greater than about 3 \Msun. The envelope temperatures in these models enter the temperature range ($> 110$ MK) where the rate of the $\iso{26}{Mg}(p,\gamma)\iso{27}{Al}$ reaction begins to exceed that of $\iso{25}{Mg}(p,\gamma)\iso{26}{Al}$. A similar trend is not seen at lower He abundances.

\subsection{Yields of neutron-capture elements}\label{sec:resultsheavyelements}
\begin{figure*}
 \begin{center}\includegraphics[width=\textwidth]{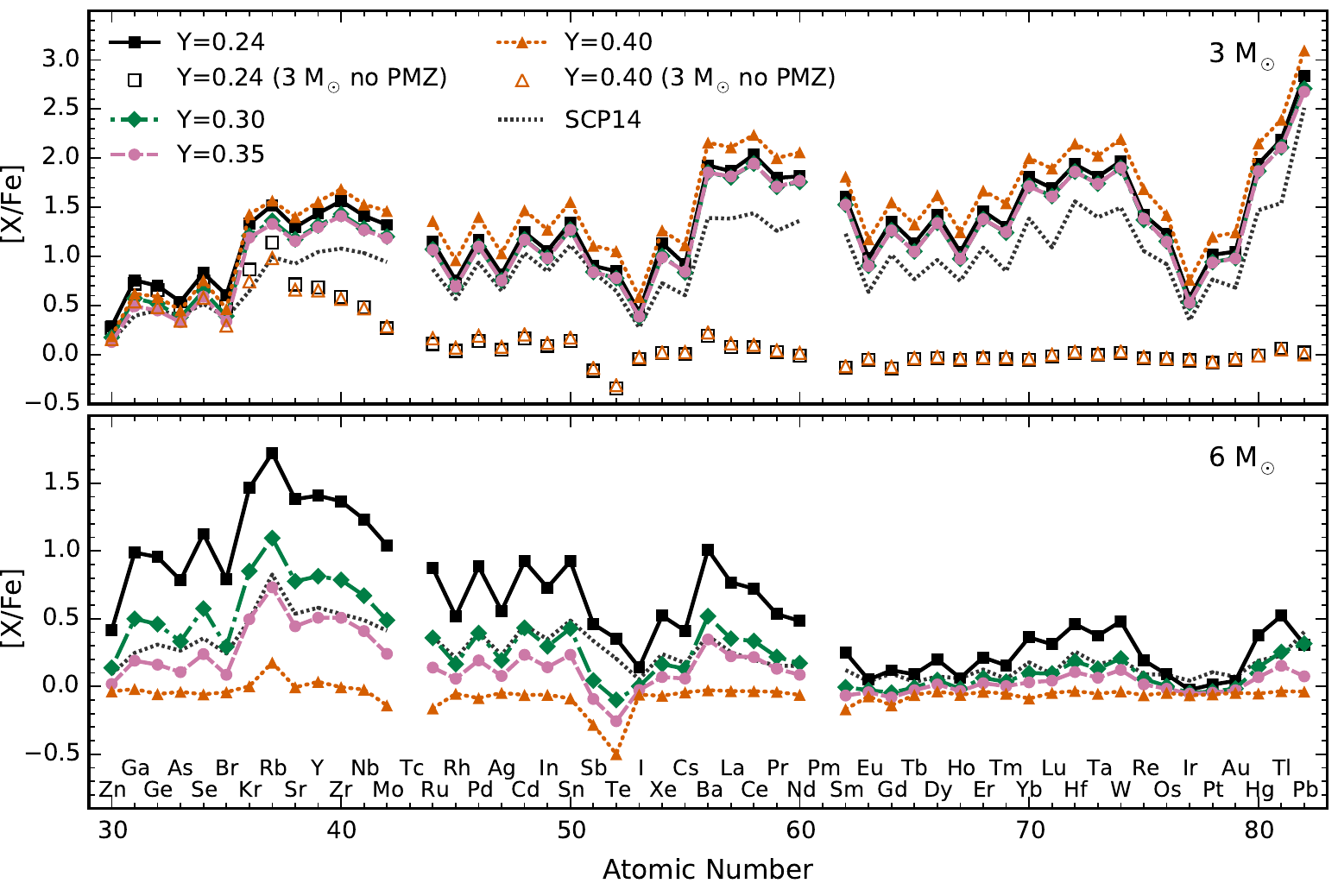}\end{center}
 \caption{Average abundances in the ejecta versus atomic number for the 3 \Msun\ models (top panel) and 6 \Msun\ models (bottom panel) for He mass fractions of $Y= 0.24$ (black squares), 0.30 (green diamonds), 0.35 (pink circles), and 0.40 (orange triangles). Open points indicate 3 \Msun\ models without a PMZ. Models with initial masses of 3 \Msun\ and below otherwise include a PMZ of $1\times10^{-3}$ \Msun.}\label{fig:yieldsm3m6}
\end{figure*}

Figure \ref{fig:yieldsm3m6} shows the heavy element abundance distributions in the average of the ejecta for the 3 \Msun\ and 6 \Msun\ models. The initial masses of 3 and 6 \Msun\ models are chosen to represent the nucleosynthetic behaviour of low- and intermediate-mass AGB models.

In the 3 \Msun\ models, the overall shape of the abundance distribution is relatively unchanged by variation to the initial He abundance. The primordial model is in reasonable agreement with the shape of the distribution predicted by SCP14, but we predict higher absolute abundances at the first and second \textit{s}-process peaks near Y and Ba by around 0.5 dex (as represented by [X/Fe] ratios). The greater abundances of these elements in our model compared with SCP14 is caused by a significantly higher TDU mass (by a factor of four) and possibly also our different approaches for modelling \iso{13}{C}-pocket formation.

In contrast with the 1.7 \Msun\ and 2.36 \Msun\ yields of KMN14 that steadily decrease with increasing He mass fraction, the Ba, La, Ce, and Pb yields of our 3 \Msun\ models exhibit a complicated dependence on the initial He content. In the 3 \Msun\ model with $Y=0.40$, the heavy-element abundances near the second \textit{s}-process peak at Ba are higher than the model with $Y=0.24$, despite the lower TDU mass. This is probably a consequence of the greater number of thermal pulses in the $Y=0.40$ model (48 versus 26 with $Y=0.24$) and a more significant effect of the \iso{22}{Ne} source.

The maximum extent of the pulse-driven convective zone ($\Mtwo{pdcz}{max}$ in Table \ref{tbl:structureagb}) gives an indication of the size of the He-rich intershell, which decreases with increasing He content. For the 3 \Msun\ models, the PMZ mass of $10^{-3}$ \Msun\  increases as a fraction of the He-intershell mass from 10 per cent with $Y=0.24$ to 33 per cent for $Y=0.40$. However, as discussed in Section \ref{sec:pmz}, the high temperatures at the base of the envelope probably prevent a significant $s$-process production from \iso{13}{C} pockets in 3 \Msun\ stars with $Y=0.40$. We expect that the model for $Y=0.40$ with no PMZ is more realistic, although there could be a very small contribution from \iso{13}{C} pockets in stars of this mass and He abundance.

For the 6 \Msun\ models, the reduction in TDU mass with He-enhancement causes heavy-element abundances in the yields to decrease monotonically with increases to the He abundance. In models with M $\gtrsim 4$ \Msun, the main neutron source is the $\iso{22}{Ne}(\alpha,n)\iso{25}{Mg}$ reaction operating in convective pulses, which produces much lower neutron-to-Fe-seed ratios and higher neutron densities compared to the \iso{13}{C} neutron source. As a result, the 6 \Msun\ abundance distributions peak at Rb, with positive [Rb/Sr] and [Rb/Zr] ratios that are characteristic of \textit{s}-process production via the \iso{22}{Ne} neutron source.

This is because the abundance ratios [Rb/Zr] and [Rb/Sr] depend on the \textit{s}-process branchings at \iso{85}{Kr} (half-life $=$ 10.8 yr) and \iso{86}{Rb} (half-life $=$ 18.6 days) \citep{vanRaai:2012fq,Karakas:2012kc}. With neutron densities above $10^8$--$10^9$ cm$^\mathrm{-3}$, neutron captures on to \iso{85}{Kr} and \iso{86}{Rb} cause the \textit{s}-process path to produce \iso{87}{Rb}. This nuclide has a magic number of neutrons, which gives it a very small neutron-capture cross section relative to \iso{85}{Rb} and neighbouring nuclides and hence it accumulates \citep{Heil:2008kl}. At high neutron densities, this causes a larger production of Rb than Sr and Zr, which are not affected by branching points \citep{DOrazi:2013bq}.

Pb and Bi represent the end point of the \textit{s}-process chain, because heavier nuclei are unstable against $\alpha$- and $\beta$-decays. Pb is not produced in significant quantities by the models that do not have a PMZ, as the low neutron-to-Fe seed ratios in these models are not sufficient to populate the \textit{s}-process chain to its heaviest nuclide.

\begin{figure*}
 \begin{center}\includegraphics[width=\textwidth]{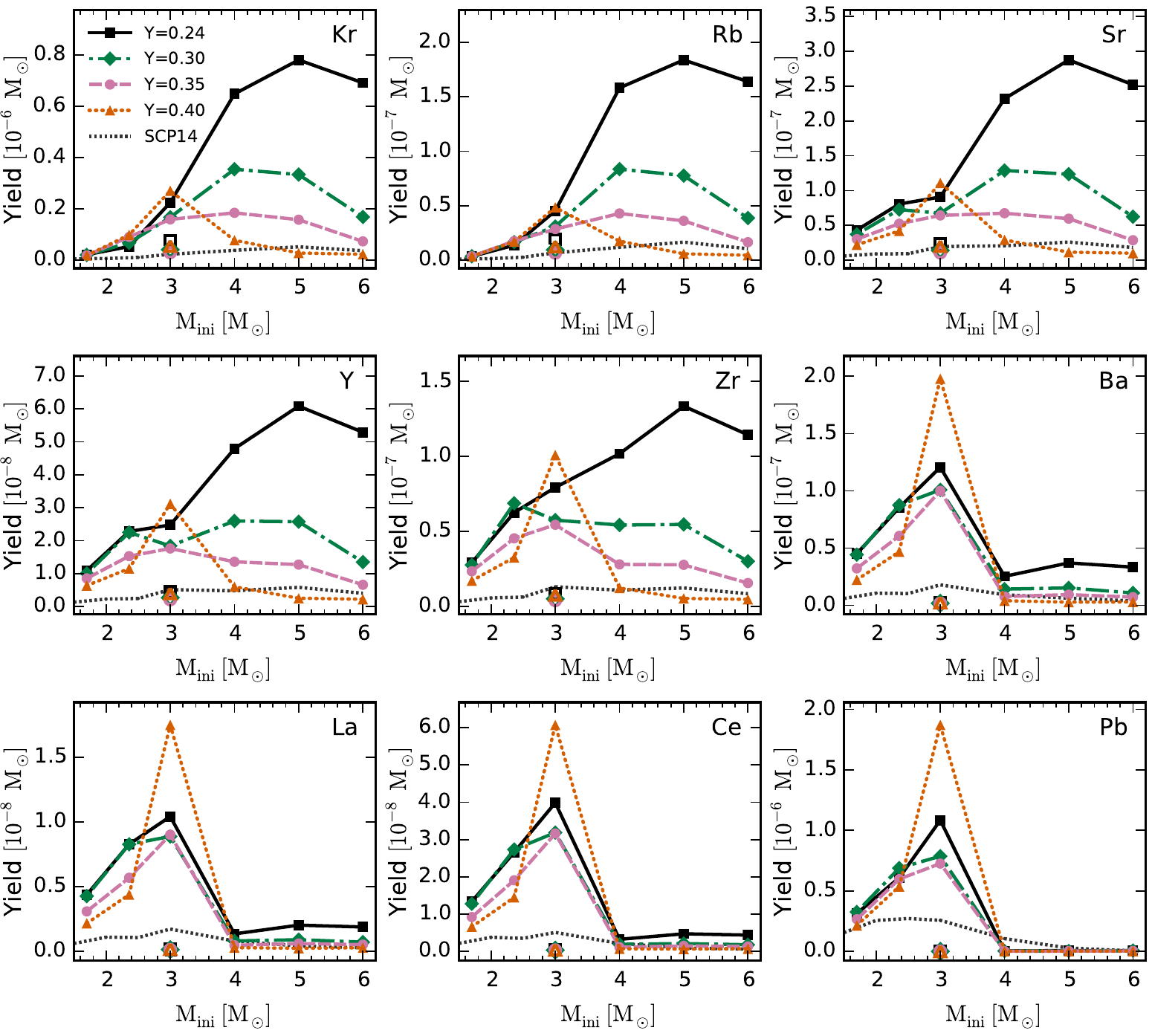}\end{center}
 \caption{Yields of selected neutron-capture elements as a function of initial mass with He mass fractions of $Y= 0.24$ (black squares), 0.30 (green diamonds), 0.35 (pink circles), and 0.40 (orange triangles). Symbols are the same as Figure \ref{fig:yieldslightelvsmass}.}\label{fig:yieldsheavyelvsmass}
\end{figure*}

Figure \ref{fig:yieldsheavyelvsmass} shows the stellar yields of selected \textit{s}-process elements as a function of initial mass for all of our models. Overall, the yields of \textit{s}-process elements exhibit a very strong reduction with He-enhancement, mostly because of the lower TDU mass.

The yields of Ba, La, Ce, and Pb increase with initial mass up to a peak near 3 \Msun. The initial rise is due to the correlation between initial mass and the amount of TDU during the AGB. The trend then reverses beyond 3 \Msun\ due to our assumption of no PMZs (and \iso{13}{C} pockets) in the higher-mass models.

For our models between 4 and 6 \Msun, in which the main neutron source is the $\iso{22}{Ne}(\alpha,n)\iso{25}{Mg}$ reaction operating in convective pulses, the elements Ba, La, Ce, and Pb are not produced in significant quantities. With He-enhancement, the yields of these elements are reduced further to effectively zero. For the lighter \textit{s}-process elements that are mainly produced by models in this mass range (Kr, Rb, Sr, Y, and Zr) the yields are decreased by an order of magnitude with $Y=0.40$.

In contrast with our results for $Y=0.24$, the SCP14 predictions for \textit{s}-process yields are substantially lower. This is connected with the smaller number of thermal pulses in their models, which result in fewer neutron producing events and TDU episodes, and possibly the different treatment of \iso{13}{C}-pocket formation. The smaller variation of the yields as a function of initial mass for the models of SCP14 reflects the flatter profile of TDU mass versus the initial mass.

\begin{figure}
 \begin{center}\includegraphics[width=\columnwidth]{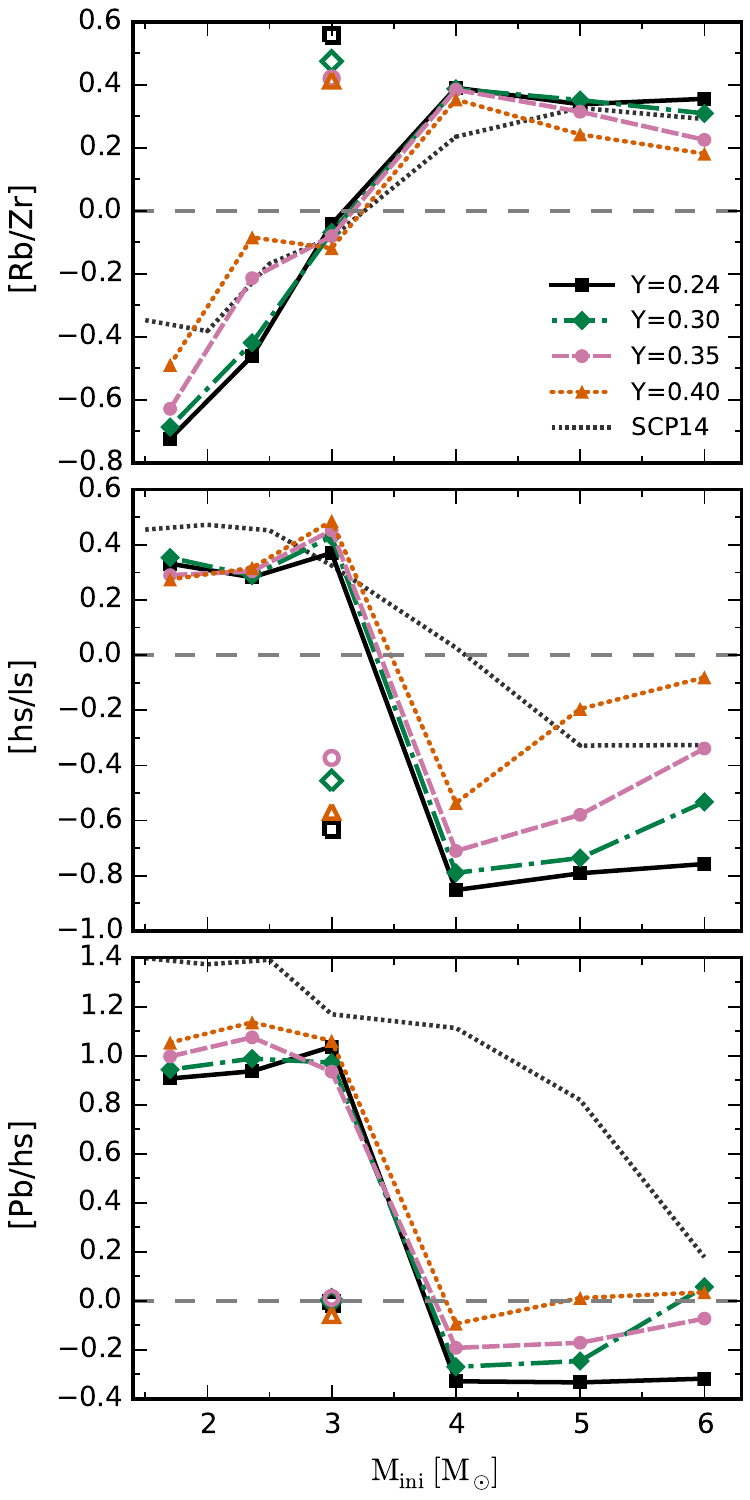}\end{center}
 \caption{The \textit{s}-process indices [Rb/Zr] (top), [hs/ls] (middle) and [Pb/hs] (bottom) of the yields as a function of initial mass with He mass fractions of $Y= 0.24$ (black squares), 0.30 (green diamonds), 0.35 (pink circles), and 0.40 (orange triangles). Symbols are the same as Figure \ref{fig:yieldslightelvsmass}.}\label{fig:yieldlshspbvsmass}
\end{figure}

The top panel of Figure \ref{fig:yieldlshspbvsmass} shows the average [Rb/Zr] ratio in the ejecta as a function of the initial stellar mass. The transition from neutrons produced in \iso{13}{C} pockets to neutrons produced by the \iso{22}{Ne} source in convective pulses (which produce higher neutron densities) at around 3 \Msun\ is marked by [Rb/Zr] values that are negative or positive, respectively. In contrast with the stellar yields of Rb and Zr (which vary by more than 1 dex between our models and SCP14), our predictions for [Rb/Zr] agree with those of SCP14 to within 0.2 dex for initial masses between 3 and 6 \Msun. This is because the [Rb/Zr] ratio is relatively independent of the TDU mass, which is substantially different in the SCP14 models.

The relative behaviour of the three \textit{s}-process peaks is quantified by the $ls$ and $hs$ indices and the Pb abundances. Following \citet{Cristallo:2011fz}, we define [ls/Fe] by
\begin{equation}
\mathrm{[ls/Fe] = ([Sr/Fe]+[Y/Fe]+[Zr/Fe])/3},
\end{equation}
and [hs/Fe] by,
\begin{equation}
\mathrm{[hs/Fe] = ([Ba/Fe]+[La/Fe]+[Nd/Fe]+[Sm/Fe])/4}.
\end{equation}

Figure \ref{fig:yieldlshspbvsmass} shows the average [hs/ls]\footnote{[hs/ls] $=$ [hs/Fe] $-$ [ls/Fe]} and [Pb/hs] indices in the ejecta of our models as a function of the initial stellar mass. Both the [hs/ls] and [Pb/hs] indices are highly sensitive to the inclusion of a PMZ (and \iso{13}{C} pockets), which results in positive values of these ratios for initial masses less than about 3 \Msun\ and negative values for higher masses.

The [hs/ls] and [Pb/hs] indices of the SCP14 models decrease more slowly with initial mass than the models presented here. This is because their treatment of the PMZ does not include an explicit cutoff at a particular initial mass. The models of SCP14 include a PMZ that gradually decreases in size and neutron production as the initial mass increases, whereas our models with masses above 3 \Msun\ do not include a PMZ.

\section{Discussion \& Conclusions}\label{sec:discussion}
We have presented and analysed the results of calculations of intermediate-mass stellar models at low metallicity with initial He mass fractions of $Y=0.24, 0.30, 0.35,$ and $0.40$.

The results of this study support the conclusion that the initial He abundance is a crucially important parameter for stellar evolution and chemical yields. For example, we have shown that the dredge up efficiency and the total TDU mass in intermediate-mass models are significantly reduced by He-enhanced initial compositions. In synthetic AGB models that require the value of $\lambda$ as a function of mass and metallicity \citep[e.g.,][]{Bertelli:2008ge,Buell:2013be}, an improvement in accuracy would be achieved if the He abundance were incorporated as a third parameter and values were obtained from full stellar evolutionary models with the appropriate initial He abundance.

One of the objectives of this study was to discover whether the reduction in the yields of Ba-peak elements found by KMN14 with He-enhanced 1.7 \Msun\ and 2.36 \Msun\ models is also the case for higher initial masses. For 4 \Msun\ and higher masses, we find that it is the case that increasing the He mass fraction results in lower yields of Ba, La, and Ce. However, our models at 4 \Msun\ and above make a small contribution to these elements with $Y=0.24$, and hence the reduction in these yields with He-enhancement will have minimal impact on chemical evolution. Of greater importance are the \textit{s}-process yields of models with M $\lesssim 3$ \Msun, which are subject to the major uncertainties of \iso{13}{C}-pocket formation. If we assume that the same PMZs form after each thermal pulse with TDU for $Y=0.40$ and $Y=0.24$, then predicted Ba-peak yields are higher by almost a factor of two, due to the near-doubling of the total number of thermal pulses in the He-enhanced model. However, the very high temperatures in the envelope during TDU and the thinner He-intershell (by a factor of three) for $Y=0.40$ suggest that a future, more advanced treatment of the PMZ would result in a much smaller production of free neutrons for the \textit{s}-process. In this case, the production of Ba-peak elements and Pb would be significantly lower at high He abundances because of the reduced contribution from \iso{13}{C} pockets.

Although many conclusions can be drawn from a grid of individual stellar yields, an understanding of their combined effect requires their application to a chemical evolution model \citep[e.g.,][]{Travaglio:1999jf,Kobayashi:2011hj,Bisterzo:2014gk,Shingles:2014ja}. From the stellar yields of the 1.7 \Msun\ and 2.36 \Msun\ models by KMN14 and the 3 to 6 \Msun\ models presented here, the chemical evolution of \textit{s}-process elements near [Fe/H] $\approx -1.4$ could be predicted as a function of He abundance, once other chemical evolution assumptions have been made \citep[e.g.,][for light and $\alpha$ elements]{Romano:2007ks}. We expect that a comparison between the chemical evolution predictions with yields of $Y=0.40$ models and the \textit{s}-process abundances in $\omega$ Centauri would be particularly insightful.

We have compared the stellar yields of our models with the primordial-He models of SCP14 and VD09, but the present lack of stellar yields for He-rich compositions makes it difficult to understand how the predicted impact of He-enhancement would be different with alternative modelling assumptions and stellar evolution codes. Of critical importance for the stellar yield predictions is the mass of intershell material that is transported to the surface by TDU. The finding of KMN14 and this work that the total mass dredged-up by TDU is highly reduced by He-enhancement (up to 96 per cent at 6 \Msun) needs to be investigated with other stellar evolution codes, which already predict much less efficient dredge-up with primordial He abundance \citep[see e.g.,][]{Mowlavi:1999ul,Lugaro:2003ew,Lugaro:2012ht}.

The increase of the core mass with initial He content is already a well-established prediction \citep[e.g.,][]{Becker:1979jk,Lattanzio:1986cz}, and therefore we expect that any predictions that are a direct consequence of larger core masses (e.g., shorter interpulse periods, shorter AGB lifetimes, and higher luminosities) would be qualitatively similar with alternative modelling assumptions. However, the predictions of AGB lifetimes and the total number of thermal pulses are highly dependent on the chosen prescription for the mass-loss rate. The increased mass-loss rate in our models with He-rich compositions would occur to a different extent with alternative mass loss prescriptions, which have a different dependence on the stellar radius and luminosity \citep[e.g.,][]{Bloecker:1995ui,Straniero:2006do}.

The changing fates of stars with the same initial mass but different He abundances leads to questions about how other mass boundaries are shifted by He-enhancement. For example, the minimum initial mass to form an electron-capture supernova at $Z=0.0006$ with primordial He abundance is about 8.2 to 8.4 $\Msun$ \citep{Doherty:2015ek}. With the larger core masses of He-enriched stars, this boundary would shift to lower masses, which are more numerous with a standard initial mass function. As a consequence, the rates of electron-capture supernovae and neutron star formation would be higher at a given star-formation rate with He-enhancement.

Similarly, He-enhancement would increase the number of hybrid CO(Ne) WDs, which have been suggested as possible progenitors of Type Iax supernovae \citep{Denissenkov:2015bi,Kromer:2015fk}. If this is the case, then the rate of Type Ia supernovae would be higher for He-rich populations.

\section*{Acknowledgments}
We thank the anonymous referee for helpful corrections to the manuscript. LJS thanks Cherie Fishlock for proofreading the manuscript. AIK was supported through an Australian Research Council Future Fellowship (FT110100475). RJS is the recipient of a Sofja Kovalevskaja Award from the Alexander von Humboldt Foundation. ML is a Momentum Project Leader of the Hungarian Academy of Sciences. This research was undertaken with the assistance of resources from the National Computational Infrastructure (NCI), which is supported by the Australian Government. This research has made use of NASA's Astrophysics Data System. The \textsc{matplotlib} package was used to generate plots \citep{Hunter:2007ih}.

\appendix

\bibliographystyle{mn2e}
\bibliography{references}

\appendix

\section{Examples of the Online Tables}

In Table \ref{atbl:elementalyields} we show the first few rows of an example online table that contains the stellar yields of each chemical element.

\begin{table*}
\begin{minipage}[]{110mm}
\caption{The first few rows of an example stellar yield table. Each table begins with a header describing the stellar model.}
\label{atbl:elementalyields}
\begin{tabular}{r r r r r r r}
\multicolumn{7}{l}{\# M$_\mathrm{ini}$ =  3.00 Msun, Z = 0.0006, Y = 0.24, M$_\mathrm{pmz}$ = 1.0E-03 Msun}\\
\multicolumn{7}{l}{\# Initial [Fe/H] = -1.41, [alpha/Fe] = 0.0}\\
\multicolumn{7}{l}{\# M$_\mathrm{remnant}$ = 0.801 Msun, M$_\mathrm{yield}$(all) = 2.199 Msun}\\
\hline
El&   Z&    log $\epsilon$(X)&[X/H]&   [X/Fe]&M$_\mathrm{yield}$&X$_\mathrm{yield}$\\
\hline
  p&   1&   12.000000&   0.000000&   1.361876&   1.441618E+00&   6.556237E-01\\
 he&   2&   11.097055&   0.167055&   1.528930&   7.210486E-01&   3.279209E-01\\
  c&   6&    9.231870&   0.761870&   2.123746&   2.961464E-02&   1.346825E-02\\
  n&   7&    7.679011&  -0.190989&   1.170886&   9.638053E-04&   4.383226E-04\\
  o&   8&    7.569763&  -1.160237&   0.201638&   8.565194E-04&   3.895308E-04\\
  f&   9&    5.327235&   0.907235&   2.269111&   5.818873E-06&   2.646327E-06\\
 ne&  10&    8.093986&   0.123986&   1.485862&   3.918737E-03&   1.782176E-03\\
 na&  11&    6.140176&  -0.099824&   1.262052&   4.578822E-05&   2.082372E-05\\
\ldots&\ldots& \ldots&     \ldots&     \ldots&         \ldots&         \ldots\\
\multicolumn{7}{l}{\texttt{\#\ \ [Rb/Zr]\ \ [ls/Fe]\ \ [hs/Fe]\ \ [hs/ls]\ \ [Pb/hs]\ \ [C+N+O/Fe]\ \ log e(CNO)}}\\
\multicolumn{7}{l}{\texttt{\ \ \ -0.2027\ \ \ 1.4595\ \ \ 1.9514\ \ \ 0.4919\ \ \ 0.9292\ \ \ \ \ \ 1.9422\ \ \ \ \ \ 9.5376}}\\
\hline
\end{tabular}
\end{minipage}
\end{table*}

\bsp

\label{lastpage}
\end{document}

%% file: table1.tex
\begin{tabular}{l r r r r r r r r r r r r}
\hline
$\M{ini}$& $Y$& $\tausub{ms}$& $\tausub{coreHe}$& $\tausub{agb}$& $\tausub{stellar}$& \M{FDU}& \M{SDU}& $\mathrm{M}^\mathrm{final}$& $\Mc^\mathrm{final}$& $\Mtwo{env}{final}$& $\Mdot^{\mathrm{final}}$& Remnant\\
& & [Myr]& [Myr]& [Myr]& [Myr]& [\Msun]& [\Msun]& [\Msun]& [\Msun]& [\Msun]& [$10^{-5}\,$\Msun]& \\
\hline
\multirow{4}{*}{3 \Msun}& 0.24& 213.2& 59.6& 0.81& 290.1& 1.79& 0.786& 1.462& 0.801& 0.661& 2.64& CO WD\\
& 0.30& 161.2& 45.5& 0.29& 219.5& 2.28& 0.844& 1.182& 0.858& 0.324& 3.06& CO WD\\
& 0.35& 126.1& 35.9& 0.23& 172.5& 2.94& 0.863& 1.628& 0.880& 0.748& 3.39& CO WD\\
& 0.40& 97.5& 29.0& 0.20& 134.8& --& 0.888& 1.641& 0.914& 0.727& 3.98& CO WD\\
\hline
\multirow{4}{*}{4 \Msun}& 0.24& 112.4& 26.6& 0.97& 146.3& --& 0.862& 1.513& 0.886& 0.626& 3.56& CO WD\\
& 0.30& 85.6& 21.4& 0.61& 112.4& --& 0.887& 1.462& 0.910& 0.552& 4.23& CO WD\\
& 0.35& 67.8& 17.5& 0.36& 89.5& --& 0.913& 1.625& 0.938& 0.687& 4.12& CO WD\\
& 0.40& 53.2& 14.5& 0.20& 71.0& --& 0.951& 1.760& 0.984& 0.776& 4.73& CO WD\\
\hline
\multirow{4}{*}{5 \Msun}& 0.24& 71.1& 14.6& 0.67& 89.3& --& 0.913& 1.671& 0.940& 0.731& 4.02& CO WD\\
& 0.30& 54.8& 11.9& 0.33& 69.4& --& 0.951& 1.597& 0.976& 0.621& 4.32& CO WD\\
& 0.35& 43.7& 10.3& 0.18& 56.0& --& 0.998& 1.603& 1.025& 0.578& 4.92& CO WD\\
& 0.40& 34.5& 8.9& 0.11& 45.0& --& 1.073& 1.911& 1.125& 0.786& 5.99& CO(Ne) WD\\
\hline
\multirow{4}{*}{6 \Msun}& 0.24& 50.0& 9.5& 0.39& 61.4& --& 0.996& 1.569& 1.021& 0.549& 4.43& CO WD\\
& 0.30& 38.8& 8.0& 0.13& 48.0& --& 1.058& 1.624& 1.080& 0.544& 5.09& CO(Ne) WD\\
& 0.35& 31.1& 6.7& 0.07& 38.8& --& 1.140& 1.632& 1.168& 0.464& 6.04& ONe WD\\
& 0.40& 24.9& 6.0& 0.04& 31.6& --& 1.234& 2.157& 1.258& 0.900& 9.15& ONe WD\\
\hline
\end{tabular}

%% file: table2.tex
\begin{tabular}{l r r r r r r r r r r r r r}
\hline
$\M{ini}$& $Y$& \Mtwo{C}{TP1}& TPs& $\left<\lambda\right>$& $\lambda_\mathrm{max}$& $\Mtdu$& $\Mtwo{pdcz}{max}$& $\Ttwo{He-shell}{max}$& $\Ttwo{H-shell}{max}$& $\Ttwo{bce}{max,dup}$& $\Ttwo{bce}{max,ip}$& $\left<\tausub{ip}\right>$& $t_\mathrm{hbb}$\\
& & [\Msun]& & & & [\Msun]& [$10^{-3}$ \Msun]& [MK]& [MK]& [MK]& [MK]& [Kyr]& [Kyr]\\
\hline
\multirow{4}{*}{3 \Msun}& 0.24& 0.786& 26& 0.94& 1.00& 0.160& 9.39& 361& 82& 57& 43& 27.71& 0\\
& 0.30& 0.847& 23& 0.86& 0.97& 0.070& 5.70& 338& 88& 63& 54& 11.61& 82\\
& 0.35& 0.867& 26& 0.79& 0.94& 0.054& 4.36& 340& 92& 64& 63& 6.87& 86\\
& 0.40& 0.892& 48& 0.70& 0.91& 0.065& 3.00& 344& 98& 78& 86& 3.63& 120\\
\hline
\multirow{4}{*}{4 \Msun}& 0.24& 0.864& 82& 0.92& 0.99& 0.281& 4.78& 350& 90& 74& 84& 9.91& 683\\
& 0.30& 0.890& 82& 0.88& 1.00& 0.186& 3.48& 354& 95& 79& 88& 5.60& 382\\
& 0.35& 0.918& 75& 0.81& 0.93& 0.114& 2.59& 355& 100& 83& 93& 3.41& 216\\
& 0.40& 0.956& 84& 0.62& 0.87& 0.058& 1.70& 349& 109& 90& 101& 1.62& 122\\
\hline
\multirow{4}{*}{5 \Msun}& 0.24& 0.915& 121& 0.90& 0.97& 0.254& 3.02& 360& 99& 82& 95& 5.08& 558\\
& 0.30& 0.955& 104& 0.84& 0.98& 0.137& 2.13& 361& 105& 89& 99& 2.85& 277\\
& 0.35& 1.002& 106& 0.70& 0.88& 0.069& 1.36& 368& 115& 97& 109& 1.38& 139\\
& 0.40& 1.077& 261& 0.26& 0.64& 0.018& 0.48& 356& 132& 118& 126& 0.34& 91\\
\hline
\multirow{4}{*}{6 \Msun}& 0.24& 0.998& 136& 0.81& 1.01& 0.140& 1.82& 384& 111& 96& 108& 2.39& 315\\
& 0.30& 1.061& 130& 0.71& 0.95& 0.058& 0.90& 381& 123& 108& 118& 0.94& 123\\
& 0.35& 1.143& 227& 0.41& 0.73& 0.021& 0.37& 382& 136& 124& 131& 0.27& 65\\
& 0.40& 1.236& 363& 0.20& 0.30& 0.005& 0.10& 353& 134& 122& 129& 0.08& 32\\
\hline
\end{tabular}